\documentclass[11pt,headings=big,numbers=noenddot,DIV=14,a4paper]{article}%

\pdfoutput=1

\usepackage[margin=1 in]{geometry}
\usepackage{comment}
\usepackage{graphicx}  
\usepackage{dcolumn} 
\usepackage{bm,relsize}   
\usepackage{amsfonts,amsmath,amssymb,amsthm,mathtools}
\usepackage{slashed}
\usepackage{placeins}
\usepackage{adjustbox}
\usepackage[normalem]{ulem}
\usepackage{lscape}
\usepackage{multirow}
\usepackage{lipsum, color}
\usepackage[usenames,dvipsnames,svgnames,table]{xcolor}
\usepackage[linktoc=page,bookmarks=false,colorlinks=true,linkbordercolor=RoyalBlue,citebordercolor=ForestGreen,urlbordercolor=CornflowerBlue]{hyperref}
\hypersetup{
    colorlinks=true,
    linkcolor=ForestGreen,
    filecolor=ForestGreen,      
    urlcolor=ForestGreen,
    citecolor=ForestGreen,
}

\usepackage[sort&compress,numbers,merge]{natbib}
\usepackage{breakcites}

\def\beq{\begin{equation}}
\def\eeq{\end{equation}}
\def\bea{\begin{eqnarray}}
\def\eea{\end{eqnarray}}

\newcommand{\GW}{\mathsmaller{\rm GW}}
\newcommand{\DM}{\mathsmaller{\rm DM}}
\newcommand{\PBH}{\mathsmaller{\rm PBH}}
\newcommand{\RH}{\mathsmaller{\rm RH}}
\newcommand{\Pl}{\mathsmaller{\rm Pl}}
\newcommand{\BL}{\mathsmaller{ B-L}}
\newcommand{\LO}{\mathsmaller{\rm LO}}
\newcommand{\NLO}{\mathsmaller{\rm NLO}}

\begin{document}

\begin{flushright}
\footnotesize
\end{flushright}
\color{black}

\begin{center}

{\LARGE \bf
%Primordial black holes from the leptogenic \\ \vspace{.2 cm} phase transition
Primordial black holes as dark matter: \\ \vspace{.4 cm} Interferometric tests of phase transition origin
}

\medskip
\bigskip\color{black}\vspace{0.5cm}

{
{\large Iason Baldes and Mar\'ia Olalla Olea-Romacho}
}
\\[1mm]

{\it Laboratoire de Physique de l'\'Ecole Normale Sup\'erieure, ENS, \\ Universit\'e PSL, CNRS, Sorbonne Universit\'e, Universit\'e Paris Cit\'e, F-75005 Paris, France}
\end{center}

\bigskip

\centerline{\bf Abstract}
\begin{quote}
We show that primordial black holes --- in the observationally allowed mass window with $f_{\PBH}=1$ --- formed from late nucleating patches in a first order phase transition imply upcoming gravitational wave interferometers will see a large stochastic background arising from the bubble collisions. As an example, we use a classically scale invariant $B-L$ model, in which the right handed neutrinos explain the neutrino masses and leptogenesis, and the dark matter consists of primordial black holes. The conclusion regarding the gravitational waves is, however, expected to hold model independently for black holes coming from such late nucleating patches.

\end{quote}

\clearpage
\noindent\makebox[\linewidth]{\rule{\textwidth}{1pt}} 
\tableofcontents
\noindent\makebox[\linewidth]{\rule{\textwidth}{1pt}}

\section{Introduction}

Putting aside the possibility being observationally misled by a theory of modified Newtonian dynamics~\cite{Milgrom:1983ca,Bekenstein:1984tv,Bekenstein:2004ne,Lelli:2016cui,Skordis:2020eui,Thomas:2023egj}, the microscopic nature of the cosmic fluid labelled dark matter (DM) is an open question. One possibility is for all of the DM to be made up of primordial black holes (PBHs)~\cite{Zeldovich:1967lct,Hawking:1971ei,Carr:1974nx}.
In this paper, we adopt the standard notation of denoting the PBH-to-DM density ratio, and assume throughout
	 \begin{equation}
	 f_{\PBH} \equiv \frac{ \rho_{\PBH} }{ \rho_{\DM } } =1.
	 \end{equation}
Observationally, this is allowed for approximately monochromatic PBH distributions with mass in the range~\cite{Carr:2021bzv}, 
	\begin{equation}
	10^{-16}M_{\odot} \lesssim M_{\PBH} \lesssim 10^{-10}M_{\odot}.
	\label{eq:windowallowed}
	\end{equation}
Here the lower limit comes from the would-be-flux of particles coming from Hawking evaporation --- which would have altered the CMB~\cite{Acharya:2020jbv,Chluba:2020oip}, produced too much extragalactic background light~\cite{Carr:2009jm,Ballesteros:2019exr,Carr:2020gox}, been detected by Voyager~\cite{Boudaud:2018hqb} or SPI/INTEGRAL~\cite{DeRocco:2019fjq,Laha:2019ssq,Laha:2020ivk} --- and the upper limit comes from microlensing observations~\cite{Niikura:2017zjd}.

In order for PBHs to be produced in the early universe, physics beyond the standard model (SM) of particles and cosmology is required. One possibility is for a feature in the inflaton potential to give an enhancement in the amplitude of overdensities at small scales~\cite{Carr:1993aq,Ivanov:1994pa}. When the overdensities re-enter the Hubble horizon, these will collapse into PBHs, provided the density contrast is sufficiently large for the given equation-of-state and shape of the curvature power spectrum~\cite{Carr:1975qj,Shibata:1999zs,Musco:2004ak,Harada:2013epa,Musco:2018rwt,Musco:2020jjb,Germani:2018jgr,Musco:2018rwt,Escriva:2019phb,Escriva:2021pmf,Escriva:2022bwe,Stamou:2023vxu}. As the evidence for inflation is rather compelling --- albeit circumstantial --- this possibility has garnered a good deal of attention. In such scenarios, gravitational waves (GWs) are produced at second order in perturbation theory, when the enhanced curvature perturbations at small scales re-enter the Hubble horizon. Such GWs may be detected at future interferometers~\cite{Nakama:2016gzw,Garcia-Bellido:2017aan,Cai:2018dig,Bartolo:2018evs,Bartolo:2018rku,Qin:2023lgo}.

An alternative is for the overdensities which collapse into PBHs to be produced at a later stage in the evolution of the universe. This could occur if there was a sufficiently strong first-order phase transition (PT) between the end of slow roll inflation and big bang nucleosynthesis. This possibility has also been the topic of a number of papers which have appeared in the last four decades, albeit more sporadically than those regarding the inflation scenario, and a number of mechanisms allowing for PBH formation during a PT have been identified~\cite{Hawking:1982ga,Crawford:1982yz,Kodama:1982sf,Hsu:1990fg,Moss:1994iq,Khlopov:1998nm,Lewicki:2019gmv,Liu:2021svg,Hashino:2021qoq,Gross:2021qgx,Baker:2021sno,Kawana:2021tde,He:2022amv,Hashino:2022tcs,Kawana:2022olo,Lewicki:2023ioy,Gouttenoire:2023naa,Salvio:2023ynn}.

Among the PT mechanisms that trigger PBH formation, the formation of large false vacuum remnants from stochastically late-nucleated patches in supercooled phase transitions stands as an intriguing possibility~\cite{Kodama:1982sf,Hsu:1990fg,Liu:2021svg,Hashino:2021qoq,He:2022amv,Hashino:2022tcs,Kawana:2022olo,Lewicki:2023ioy,Gouttenoire:2023naa,Salvio:2023ynn}. This is an unavoidable process if sufficiently strong PTs take place. The remnants lead to Hubble sized overdensities being present in the radiation following post-PT reheating, most closely emulating the inflation mechanism. The concurrent production of smaller PBHs has also recently been considered in~\cite{Lewicki:2023ioy}, using a Hoop conjecture criterion, but the spectrum remains relatively peaked.\footnote{For PBHs produced from PTs during or at the end of inflation see~\cite{Garriga:2015fdk,Deng:2016vzb,Deng:2017uwc,Kusenko:2020pcg,Ashoorioon:2020hln,Animali:2022otk}.}

The reheating following bubble percolation in the PT leads to an initially inhomogenous state, not modelled in General Relativistic numerical simulations of PBH formation~\cite{Shibata:1999zs,Musco:2004ak,Harada:2013epa,Musco:2018rwt,Musco:2020jjb,Germani:2018jgr,Musco:2018rwt,Escriva:2019phb,Escriva:2021pmf,Escriva:2022bwe}, but the large number of bubbles present and the fact that an $\mathcal{O}(1)$ overdensity is achieved, leads us to assume that collapse is rather plausible. Therefore, for the scope of this paper, we operate under the assumption that this mechanism functions as expected, along the lines of~\cite{Liu:2021svg,Hashino:2021qoq,He:2022amv,Hashino:2022tcs,Kawana:2022olo,Gouttenoire:2023naa,Salvio:2023ynn}. Nevertheless, we warn the reader that the collapse criterion in this scenario, which will necessarily feature a certain level of anisotropy, turbulent inhomogeneity, non-sphericity, and initial velocity vectors for the radiation fluid differing from the inflationary overdensity scenario, is far from certain.

Strong first order PTs also produce a stochastic background of GWs~\cite{Witten:1984rs,Hogan:1986qda}. The production of the PBHs in the PT will lead to GWs from the bubble collisions, and these GWs are observable in upcoming interferometers, as we explain next using a simple argument. In the rest of the subsequent paper we provide a more detailed calculation.

\section{A simple argument}

In the delayed patch mechanism, the mass of the PBH is set by the energy contained inside a Hubble volume during the PT,
	\begin{equation}
	M_{\PBH} \sim \frac{M_{\Pl}^{3}}{T_{\RH}^2},
	\label{eq:PBHmassapprox}
	\end{equation}
where $M_{\rm Pl} \sim 10^{19}$ GeV is the Planck mass, and $T_{\rm RH}$ is the reheating temperature following the completion of the PT (we assume rapid reheating). The observationally allowed mass window for which $f_{\PBH}=1$ is possible, Eq.~\eqref{eq:windowallowed}, constrains the reheating temperature to lie between~\cite{Gouttenoire:2023naa}
	\begin{equation}
	10 \; \mathrm{TeV} \lesssim T_{\RH} \lesssim 10^{4} \; \mathrm{TeV} .
	\end{equation}
The abundance is controlled by the time the PT takes to complete compared to the Hubble rate. To achieve an abundance of $f_{\PBH}=1$, we require the inverse timescale of the transition to be~\cite{Gouttenoire:2023naa},
	\begin{equation}
	\beta \equiv \frac{1}{\Gamma_{\rm bub}}\frac{d\Gamma_{\rm bub}}{dt} \approx 8H,
	\end{equation}
where $\Gamma_{\rm bub}$ is the bubble nucleation rate per unit volume and $H$ is the Hubble rate during the PT.

The amplitude of GWs from bubble collisions during the PT scales as $\propto R^{2} \sim v_{\rm w}^2/\beta^{2}$, where $R$ is the bubble radius and $v_{w}$ is the wall velocity. For the PTs of interest here, it is a good approximation to set $v_{w} \simeq 1$, and we do so throughout the rest of this paper. The precise amplitude and spectral shape of the GWs has been addressed in a number of works~\cite{Kosowsky:1992rz,Kosowsky:1991ua,Kosowsky:1992vn,Kamionkowski:1993fg,Caprini:2007xq,Huber:2008hg,Jinno:2017fby,Konstandin:2017sat,Cutting:2018tjt,Cutting:2020nla,Lewicki:2020jiv,Lewicki:2020azd}. In our calculations we shall use the recent determinations, relevant for our type of PTs, in which the peak amplitude redshifted to today is~\cite{Konstandin:2017sat,Cutting:2020nla,Lewicki:2020azd}
	\begin{equation}
	\Omega_{\GW} \sim 10^{-6} \left( \frac{H}{\beta} \right)^{2}.	
	\end{equation}
Said GW estimates still feature uncertainties for the very strong PTs we will be interested in, due to the expansion of the universe during the PT itself, see~\cite{Zhong:2021hgo,Bringmann:2023opz}. So in this regard our results should be considered preliminary, albeit promising, until the GW predictions can be further refined. 
The peak frequency of the GW signal is similarly set by the inverse bubble radius at the collision time, which is then redshifted to today,
	\begin{equation}
	f_{\rm peak} \sim 1 \; \mathrm{mHz}  \left( \frac{\beta}{H} \right) \left( \frac{T_{\RH}}{ 100 \; \mathrm{TeV} } \right).
	\label{eq:freqapprox}
	\end{equation}
From the above, we see that in such a $f_{\PBH}=1$ from a PT scenario, we will have a peak amplitude $\Omega_{\GW} \sim 10^{-8}$, i.e.~well above stochastic astrophysical foregrounds, somewhere in the frequency range $10^{-3} \; \mathrm{Hz} \lesssim f_{\rm peak} \lesssim 1 \; \mathrm{Hz}$. The GWs at lower peak frequencies are detectable by LISA~\cite{amaroseoane2017laser}. Those at higher peak frequencies are similarly detectable by the Einstein Telescope (ET)~\cite{Maggiore:2019uih}. A mid-frequency interferometer such as BDECIGO~\cite{Kawamura:2020pcg} would be sensitive to the entire allowed PBH mass range and could in any case help confirm and characterise the signal. 

Similar conclusions hold if we replace a detector with a variant of broadly similar equivalence, such as ET with Cosmic Explorer~\cite{Evans:2021gyd} or BDECIGO with AEDGE~\cite{AEDGE:2019nxb}.  Note we consider only the GWs from the bubble wall collisions, the calculation and addition of GWs sourced from the enhanced curvature perturbations at small scales in the context of the PT mechanism~\cite{Liu:2022lvz}, is left for future work.

The rest of the paper is dedicated to calculating the PBH production and GW signal within a simple beyond the SM scenario, motivated by neutrino masses and leptogenesis, for which the PBHs will act as a DM candidate. We will scan over the parameters giving $M_{\PBH}$ in the allowed window. The model we choose is a classically scale invariant, gauged $B-L$, extension of the SM. Such close-to-conformal models can provide the necessary supercooled PTs~\cite{Jinno:2016knw,Hambye:2018qjv,Marzo:2018nov}. Even though fine-tuning is required to get the cosmological constant close-to-zero today, this seems to be the case almost generically when dealing with the vacuum energy density in a cosmological context. We therefore put questions of fine-tuning aside, which may anyway similarly be present in other models triggering PBH formation, and simply explore the scenario of achieving the observed $\rho_{\DM}$.

We find it beneficial to conduct an explicit calculation within the aforementioned beyond the SM physics scenario. The calculation in such a model serves to illuminate several facets of the subject matter. For instance, many previous studies (e.g.~\cite{Liu:2021svg,Gouttenoire:2023naa}) assume a model independent approximation for the bubble nucleation rate
	\begin{equation}
	\Gamma_{\rm bub}(t) = H^{4}(t_n)e^{\beta (t - t_{n})},
	\label{eq:gammaapprox}
	\end{equation}
where $t_{n}$ is the bubble nucleation time, by definition when $\Gamma_{\rm bub} = H^{4}$, and $\beta$ acts as the coefficient of the first order Taylor approximation of the full nucleation rate. One may wonder, however, whether the second and possibly higher derivatives will effectively play a role in setting $f_{\PBH}$, because the probability of a Hubble patch collapse is very sensitive to the precise behaviour of $\Gamma_{\rm bub}$. Nevertheless, our results below show the approximate form yields accurate predictions for the required $\beta$, at least in close-to-conformal potentials. Thus confirming the applicability of the GW signal based on the requirements on $\beta$ for achieving $f_{\PBH}=1$, as outlined in our discussion around Eqs.~\eqref{eq:PBHmassapprox}-\eqref{eq:freqapprox}.

Let us now also mention some work which previously headed along this path, but with different overall scope, e.g.~\cite{Liu:2021svg,Hashino:2021qoq,Hashino:2022tcs,Xie:2023cwi,Banerjee:2023brn}.\footnote{As this paper was in the final stages of preparation, an analytic approach to estimating the PT parameters, GW and PBH production, partly in the context of the $B-L$ model, was presented in~\cite{Salvio:2023ynn}.} Indeed, already in 1984, Witten was discussing the concomitant aspects of a strong QCD phase transition, GWs, production of PBHs, and future pulsar timing array measurements~\cite{Witten:1984rs}. Compared to the aforementioned papers, we provide a thorough scan over the expected PBH mass range coming from the PT, giving a more detailed picture of the expected GW signal. We also further develop the late nucleating patch formalism~\cite{Liu:2021svg,Gouttenoire:2023naa}, showing the calculation can be recast from using time as a variable to a conveniently chosen temperature, here that of the false vacuum plasma.

Our focus here is on $f_{\PBH}=1$ which results in a scale of new physics beyond collider reach. The interplay of micro-lensing hints for $0 <  f_{\PBH} < 1$ at larger PBH masses, GW signals from supercooled PTs, and collider physics has instead recently been studied in~\cite{Gouttenoire:2023pxh}.

\section{The Model}

\subsection{Particles and charges}

As our example, we take a classically conformal $B-L$ extension of the SM, in which we add three right handed neutrinos, $N_{i}$ with $Q_{\BL}=-1$, to cancel off anomalies~\cite{Iso:2009ss,Iso:2009nw,Jinno:2016knw}. To break the $B-L$ symmetry, we add a complex scalar, $\rho$, with $Q_{\BL}=-2$. Ultimately, the radiative breaking of the $B-L$ symmetry induces a negative mass term for the electroweak Higgs field via a portal coupling $\lambda_{\rho h}$, thus eventually breaking electroweak symmetry. The kinetic term for the new scalar reads
	\begin{equation}
	\mathcal{L} \supset (D_{\mu}\rho)^{\ast}(D^{\mu}\rho),
	\end{equation}
where $D_{\mu} \equiv \partial_{\mu} - 2ig_{\BL}Z'_{\mu}$ is the covariant derivative with gauge coupling $g_{\BL}$. This will generate a mass for the gauge boson once $\rho$ gains a vacuum expectation value (vev) radiatively, $\langle \rho \rangle = v_{\rho}/\sqrt{2}$. At the classical level, the scalar potential reads
	\begin{equation}
	V(H,\rho) = \lambda_{\rho} |\rho|^4 + \lambda_{ \rho  h}  |\rho|^2 |H|^2 + \lambda_{h } |H|^4, 
	\end{equation}
where $H$ is the SM Higgs doublet. In addition to the SM Yukawas we also have
	\begin{equation}
	\label{eq:yuks}
	\mathcal{L} \supset y_{\nu i j} \overline{l_{Li}} \tilde{H} N_{j} +  \frac{1}{2}y_{Ni} \rho \overline{N} N^c + \mathrm{H.c.},
	\end{equation}
where $l_{Li}$ are the SM lepton doublets and $\tilde{H} \equiv i \sigma_{2} H^{\ast}$. The second term in Eq.~\eqref{eq:yuks} will give Majorana masses to the $N_{i}$. These can then decay in a CP violating manner via the first Yukawa coupling realising leptogenesis, on which we will briefly comment later. The neutrino oscillation data~\cite{Davis:1968cp,SNO:2001kpb,Super-Kamiokande:1998kpq,K2K:2006yov,MINOS:2011qho,DayaBay:2012fng,RENO:2012mkc,Esteban:2020cvm} is explained through a type-I seesaw~\cite{Minkowski:1977sc,Yanagida:1979as,Gell-Mann:1979vob,Mohapatra:1979ia} (here at relatively low scales $\lesssim 10^{8}$ GeV). In the regime we are interested in, there is a significant hierarchy between $v_{\rho}$ and the electroweak vev $v_{\rm ew} \simeq 246$ GeV, which implies a small cross quartic coupling,
	\begin{equation}
	\lambda_{\rho h} \simeq -\left( \frac{m_{h}}{v_{\rho}} \right)^{2},
	\end{equation} 
where $m_{h} \simeq 125$ GeV is the SM-like Higgs mass. This hierarchy situates the scenario beyond the reach of current collider constraints.

\subsection{Radiative symmetry breaking at finite temperature}

\begin{figure}[t]
\begin{center}
\includegraphics[width=220pt]{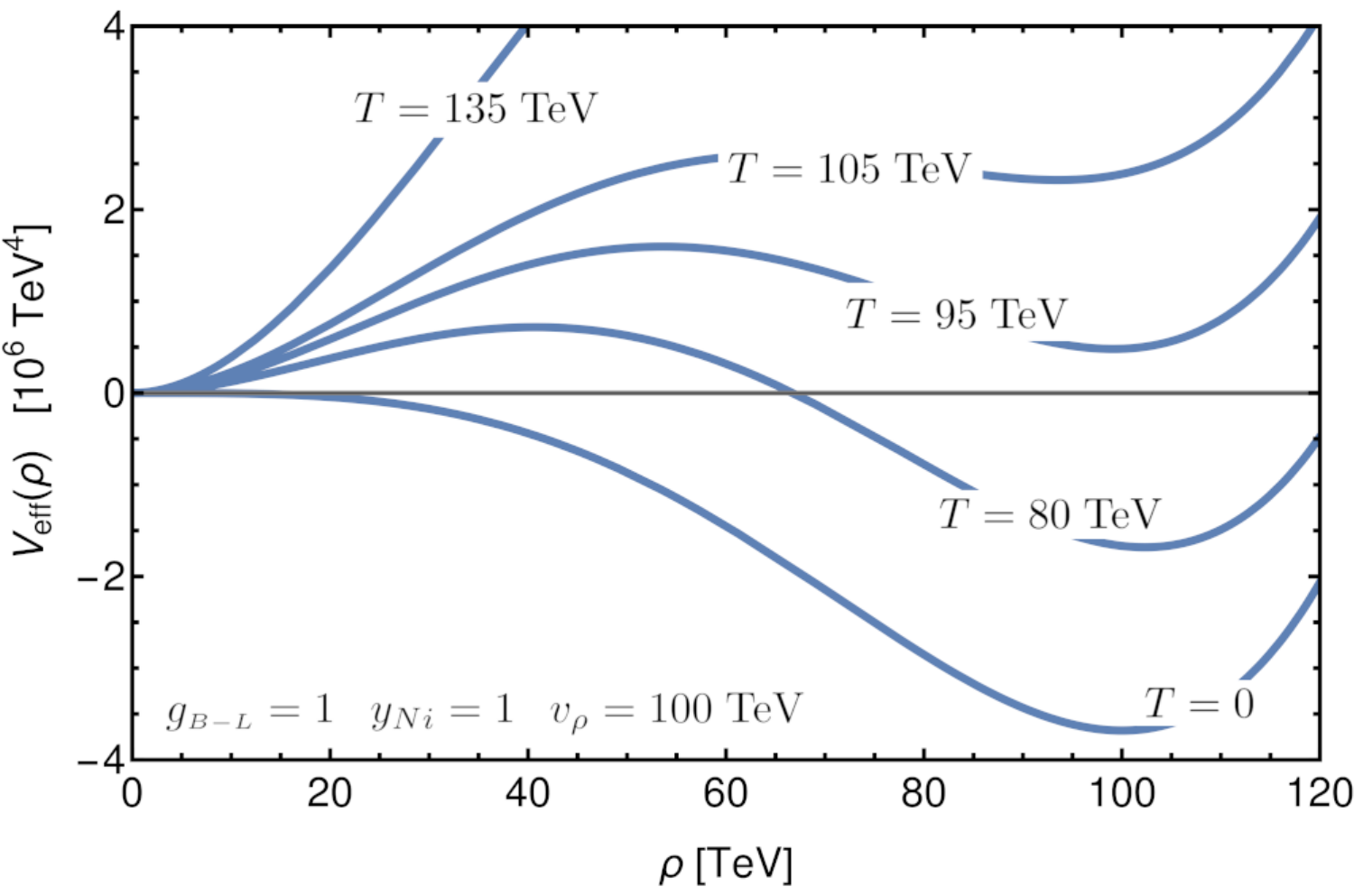}
\end{center}
\caption{\small The evolution with temperature of the effective potential for a benchmark point with $g_\BL=1$, $y_{Ni}=1$ and $v_\rho=100 \; \mathrm{TeV}$. The barrier separating the two minima means the evolution from the false to the true vacuum corresponds to a first order phase transition. The barrier is always present when the finite temperature corrections are added to the radiatively induced symmetry breaking potential.}
\label{fig:potential}
\end{figure}

Our treatment of the field theory in this section, leading to the approximation of the effective potential, $V_{\rm eff}$, follows~\cite{Hambye:2018qjv}. At one loop the beta function of the $B-L$ Higgs quartic is 
	\begin{equation}
	\beta_{\lambda_\rho} \equiv \frac{ d \lambda_{\rho} }{ d \log \mu } = \frac{1}{(4\pi)^2} \left( 96 g_{B-L}^{4}- y_{Ni}^{2} + 2 \lambda_{\rho h}^2 + 20 \lambda_{\rho}^{2} + \lambda_{\rho} [ 2 y_{Ni}^2 - 48 g_{B-L}^{2} ]  \right),
	\end{equation} 
where the sum over the $y_{Ni}$ is understood ($i=1,2,3$). 
Consistency requires $\beta_{\lambda_\rho} > 0$, so that $\lambda_{\rho}$ turns negative when running down from high scales, triggering radiative symmetry breaking. Making the here justifiable approximation of the symmetry breaking as dominantly in a single field direction, following the approach of Gildener and Weinberg~\cite{Gildener:1976ih}, we write the one loop zero temperature potential as
	\begin{equation}
	V_0(\rho) = \beta_{\lambda_\rho} \frac{ \rho^{4} }{ 4 } \left( \log \left[ \frac{ \rho } { v_{\rho} } \right] - \frac{1}{4} \right),
	\end{equation}
where from now we denote $v_{\rho}$ as the vev at the zero-temperature minimum and $\rho$ denotes the classical field value. At the $B-L$ symmetry breaking scale, $ \lambda_{\rho} \simeq 0$, because the coupling is switching signs. We also require a small $\lambda_{\rho h}$, as mentioned above. Therefore we can approximate the beta function as
	\begin{equation}
	\beta_{\lambda_\rho} \approx \frac{1}{(4\pi)^2} \left( 96 g_{B-L}^{4}- y_{Ni}^{2} \right).
	\end{equation}
The mass of the physical scalar after symmetry breaking is $m_{\rho} \simeq \sqrt{\beta_{\lambda}}v_{\rho}$. Its field dependent mass gives only a small contribution to the thermal corrections in the effective potential.
The dominant thermal corrections instead come from the field dependent mass of the $B-L$ gauge boson, 
	\begin{equation}
	M_{Z'} = 2 g_{\BL} \rho .
	\end{equation}
and the heavy Majorana neutrinos,
	\begin{equation}
	M_{Ni} = \frac{ y_{Ni} \rho }{\sqrt{2}}.
	\end{equation}
Their thermal contributions are
	\begin{equation}
	V_{T}(\rho,T) = \frac{T^{4} }{ 2\pi^{2} }\left( 3 J_{B}\left[ \frac{ M_{Z'}^2}{T^2} \right] + 2 J_{F}\left[ \frac{ M_{Ni}^2 }{ T^2 } \right] \right),
	\end{equation}
where the thermal functions are
	\begin{equation}
	J_{B/F}(x) = \pm \int_0^\infty dk \; k^2 \, \log[1 \mp e^{-\sqrt{k^2+x}}],
	\end{equation}
for bosons and fermions respectively. To improve the perturbative analysis, we include the corrections due to the resummation of daisy diagrams~\cite{Arnold:1992rz},
	\begin{equation}
	V_{\rm daisy}(\rho,T) = \frac{T}{12\pi} \left( M_{Z'}^3 - [M_{Z'}^2 +  \Pi_{\rm Z'} ]^{3/2} \right),
	\end{equation}
where the thermal mass of the longitudinal component of the $B-L$ gauge boson is
	\begin{equation}
	\Pi_{\rm Z'} = 4 g_{B-L}^{2} T^{2}.
	\end{equation}
The full effective potential we consider is therefore
	\begin{equation}
	V_{\rm eff}(\rho,T) = V_0(\rho) + V_{T}(\rho,T) + V_{\rm daisy}(\rho,T),
	\end{equation}
after the aforementioned approximations have been made. To be consistent with observations, showing a small vacuum energy in the present day, a cosmological constant term,
	\begin{equation}
	\Lambda_{\rm vac} = \frac{ \beta_{\lambda_\rho} v_{\rho}^{4} }{ 16 }
	\end{equation}
needs to be added to the potential, so that $V_0(v_{\rho}) = 0$. The gravitationally important terms, $\Lambda_{\rm vac}$, together with the Planck mass, $M_{\rm Pl}$, must be considered external to our classically scale invariant sector. An example showing the temperature evolution of the potential is shown in Fig.~\ref{fig:potential}. Nucleation is possible once the false and true minima become degenerate at the critical temperature (time), $T_{c}$ ($t_{c}$).

\subsection{Bubble nucleation and percolation}

The bubble nucleation rate per unit volume can be approximated as~\cite{Coleman:1977py,Callan:1977pt,Linde:1981zj}
	\begin{equation}
	\Gamma_{\rm bub} = A^{4} e^{-S},
	\end{equation}
where $T$ is the temperature, $A$ is a mass dimension one pre-factor we elucidate later, and $S$ is the Euclidean bounce action. At zero temperature the configuration minimizing the action, $S \equiv S_{4} $, is $O(4)$ symmetric and
	\begin{equation}
	 S_{4}=2\pi^2\int_{0}^{\infty} dr~r^3~\left(\frac{1}{2}\left[\frac{ d \rho }{dr} \right]^2+V_{\rm eff}\left(\rho\right)\right)
	 \end{equation}
In this work we can limit ourselves PT dominated by single field directions. The action $S_{4}$ corresponds to quantum tunneling through the potential barrier. The equation of motion is
 	\begin{equation}
	\frac{ d^2 \rho }{dr^2} + \frac{3}{r}\frac{ d \rho }{dr} = \frac{dV_{\rm eff}}{d\rho},
	\end{equation}
with boundary conditions
	\begin{equation}
	\frac{ d \rho }{dr}\Big|_{r=0} = 0, \qquad \textrm{and} \qquad  \lim_{r \to \infty} \rho(r) = 0.
	\end{equation} 
At finite temperature the field instead becomes periodic in $1/T$ in the time coordinate. The configuration minimizing the action is $O(3)$ symmetric. Furthermore, at sufficiently high temperatures, the minimum action configuration becomes constant in the time direction and
	\begin{equation}
	S \equiv \frac{ S_{3} }{ T } = \frac{ 4\pi }{T}\int_{0}^{\infty} dr~r^2~\left(\frac{1}{2}\left[\frac{ d \rho }{dr} \right]^2+V_{\rm eff}\left(\rho,T\right)\right).
	\end{equation}
This represents bubble formation through classical field excitation over the barrier. The corresponding equation of motion is then
 	\begin{equation}
	\frac{ d^2 \rho }{dr^2} + \frac{2}{r}\frac{ d \rho }{dr} = \frac{dV_{\rm eff}}{d\rho},
	\end{equation}
with the same boundary conditions as above. The solution with a non-trivial periodic bounce in the time coordinate, corresponding to quantum tunneling at finite temperature, is more difficult to evaluate in practical calculations. It is eventually, however, well approximated at low $T$ by $S_{4}$ but calculated with $V_{\rm eff}$ including any finite temperature corrections. For simplicity, we therefore take a rather standard estimate for our nucleation rate
	\begin{equation}
	\Gamma_{\rm bub} \approx {\rm Max} \left[   \frac{1}{R_c^{4}} \left( \frac{S_4}{2\pi} \right)^2 e^{-S_4}, \quad T^4\left( \frac{S_3}{2\pi T} \right)^{3/2} e^{-S_3/T}   \right],
	\end{equation}
where $R_c\sim 1/T$ is the bubble radius in the low $T$ limit of radiative symmetry breaking, and we have used standard estimates of the pre-factor (for discussion regarding uncertainties see~\cite{Croon:2020cgk}). For the PTs we study, we have verified that $\Gamma_{\rm bub}$ is given by the $S_{3}$ action.

The nucleation temperature in an average Hubble patch, $T_n$, or the equivalent nucleation time, $t_{n}$, is defined as when the nucleation rate equals the Hubble rate,
	\begin{equation}
	\Gamma_{\rm bub} = H^{4}.
	\end{equation}
In practice, we compute the nucleation temperature using the Hubble rate in the false vacuum,
	\begin{equation}
	 H \equiv H_{\rm false} = \sqrt{ \frac{8 \pi}{3 M_{\rm Pl}^2 } \left( \frac{ g_{\ast} \pi^{2} }{ 30 } T^{4} + \Lambda_{\rm vac} \right)} ,
	\end{equation}
where $T$ is the false vacuum temperature, and $g_{\ast}$ are the radiation degrees-of-freedom. In our scenario, $g_{\ast} \simeq 116$ prior to the PT, and $g_{\ast} \simeq 113$ just after the PT. Henceforth, any instances of the Hubble parameter and the scale factor presented without an index should be understood as these quantities evaluated in the false vacuum phase. We will be interested in strong supercooled phase transitions for which the universe becomes vacuum dominated. This occurs at a temperature
	\begin{equation}
	T_{\rm infl} =  \left( \frac{30 \Lambda_{\rm vac} }{g_{\ast}(T_{\rm infl})\pi^{2}}  \right)^{1/4},
	\end{equation}
For the case we will be interested in, namely efficient reheating and negligible changes in degrees-of-freedom between the two phases, we have $T_{\RH} \simeq \mathrm{Max}[T_{\rm infl},T_p]$, where $T_{\RH}$ is the reheating temperature following the completion of the phase transition, and $T_p$ is the bubble percolation temperature, which we now review how to calculate.

\begin{figure}[t]
\begin{center}
\includegraphics[width=200pt]{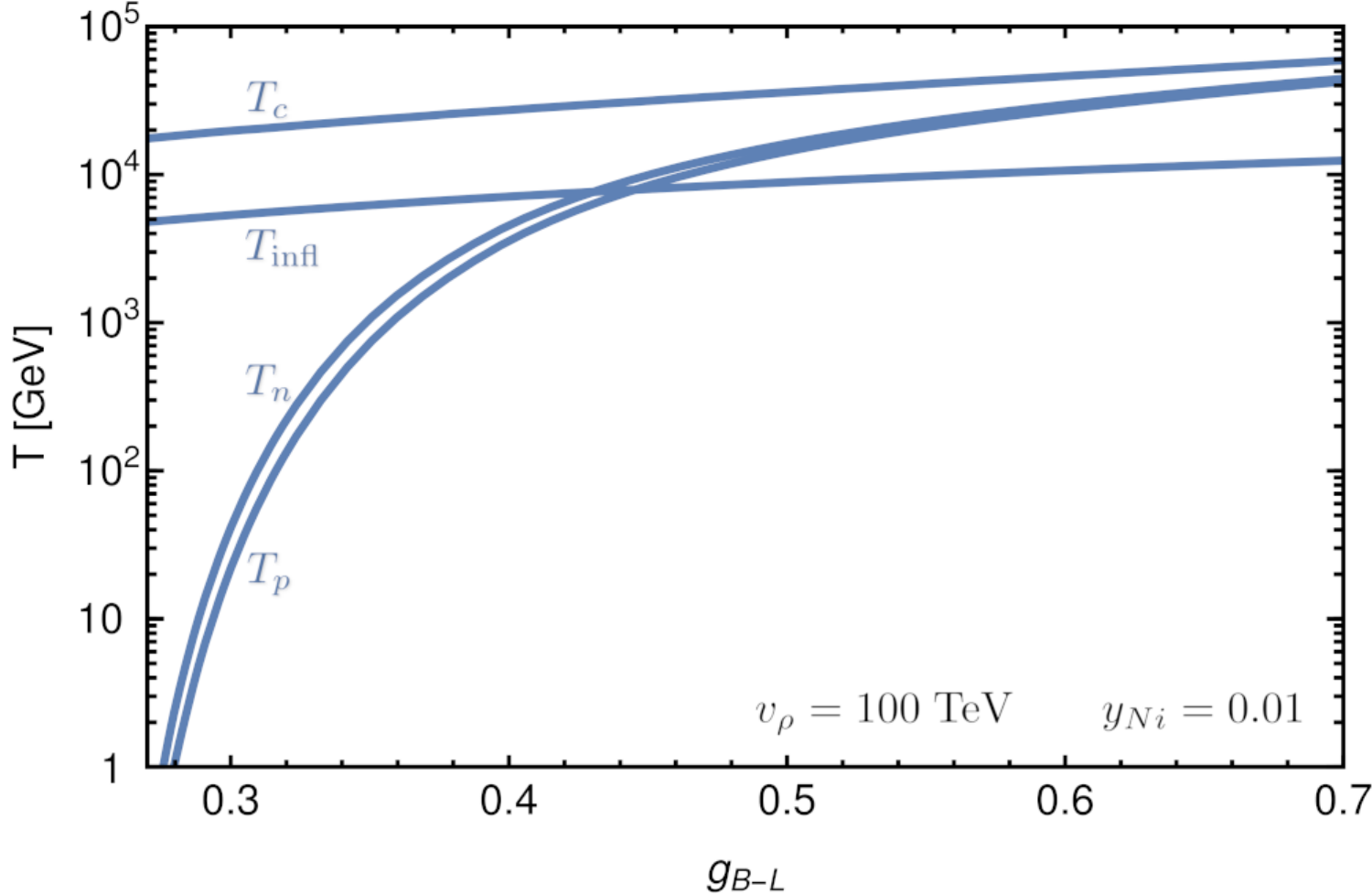} $\quad$
$\quad$
\includegraphics[width=200pt]{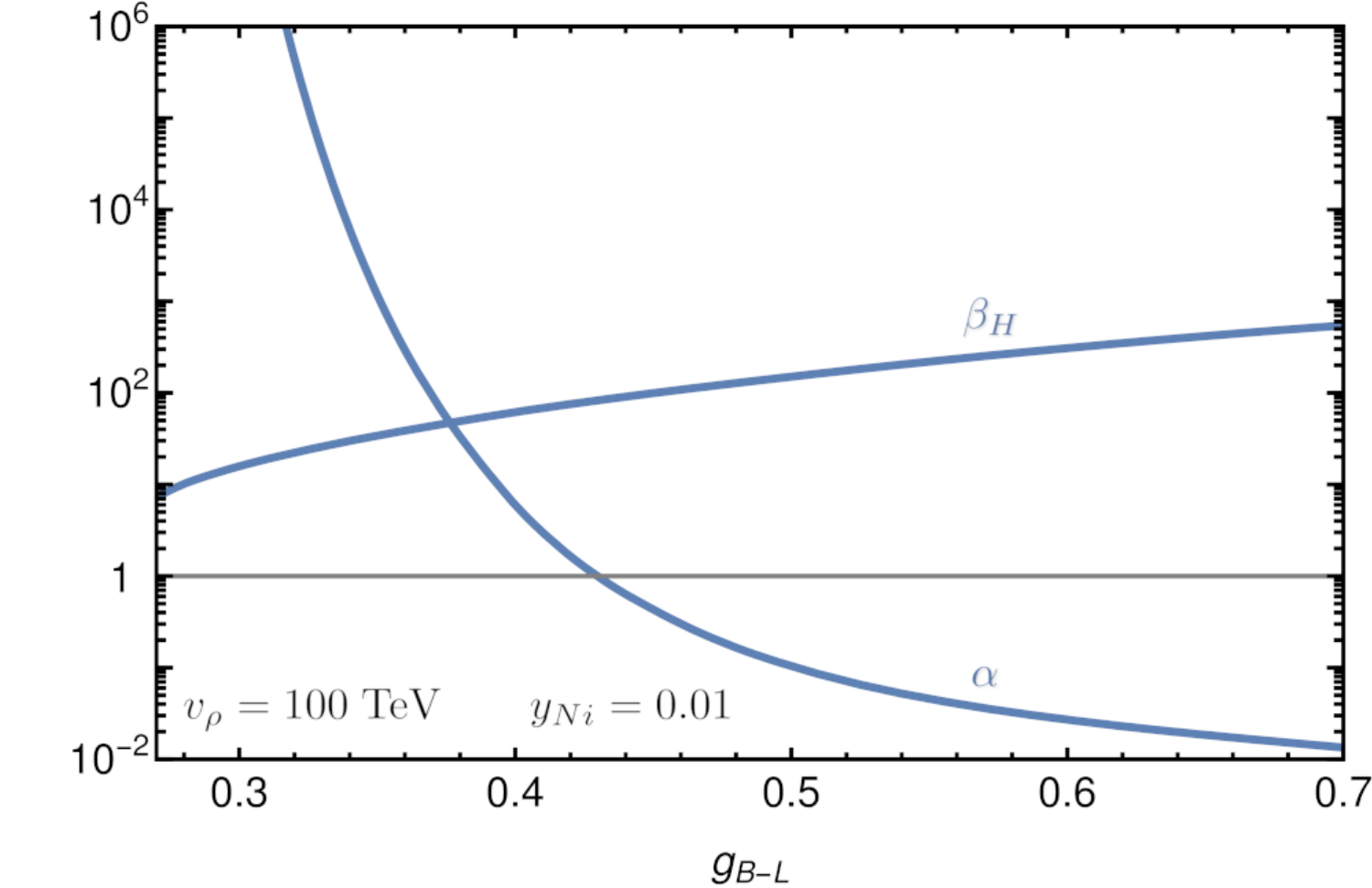}
\end{center}
\caption{\small Left: Temperatures characterising the phase transition as a function of the gauge coupling. We display the following temperatures: critical, $T_{c}$, nucleation , $T_{n}$, percolation, $T_{p}$, and the temperature at which thermal inflation starts, $T_{\rm infl}$. Right: The dimensionless phase transition latent heat, $\alpha$, and inverse timescale normalised to the Hubble rate, $\beta_{H}$.}
\label{fig:PTexample}
\end{figure}

In the vacuum dominated regime, the temperature (time) at which the bubbles percolate, $T_p$ ($t_p$), may be appreciably different from $T_n$ ($t_n$). It is useful to introduce the comoving radius of a bubble at false vacuum temperature $T$, nucleated at some higher temperature $T'$,
	\begin{equation}
	r(t,t') = \int_{t'}^{t}\frac{d\tilde{t}}{a(\tilde{t})}  = r(T,T')  = \int_{T}^{T'} \frac{d\tilde{T}}{\tilde{T}H(\tilde{T})a(\tilde{T})},
	\label{eq:comovingrad}
	\end{equation} 
where $a$ is the scale factor, we have assumed $v_{w} \simeq 1$ shortly after nucleation, and have used $dT = -T H dt$. The physical radius is then 
	\begin{equation}
	R(T,T') = a(T) r(T,T') = a(T) \int_{T}^{T'} \frac{d\tilde{T}}{\tilde{T}H(\tilde{T})a(\tilde{T})} = \frac{1}{T}\int_{T}^{T'} \frac{d\tilde{T}}{H(\tilde{T})},
	\label{eq:physbubsize}
	\end{equation}
where we have assumed a constant $g_{\ast}$ and so used $a(T) T = a(\tilde{T}) \tilde{T} $. The expected volume of true-vacuum bubbles per comoving volume (double counting overlapping regions and including fictitious nucleations in true vacuum) is given by~\cite{Guth:1979bh,Guth:1981uk,Guth:1982pn,Enqvist:1991xw,Turner:1992tz}
	\begin{align}
	I   = \frac{ 4\pi }{ 3 } \int_{t_c}^{t} dt' \Gamma_{\rm bub}(t') a^3(t')r^3(t,t') = \frac{4\pi}{3}\int_{T}^{T_c}dT'\frac{ \Gamma_{\rm bub}(T') }{ T'^{4}H(T') }\left(\int_{T}^{T'} \frac{ d\tilde{T} }{ H(\tilde{T}) } \right)^{3}.
	\label{eq:Ifunc}
	\end{align}
To find the probability of a given point in the comoving volume to be in the false vacuum, we need to exclude fictitious nucleations and avoid double counting overlapped regions, and the appropriate expression is given by~\cite{Guth:1979bh,Guth:1981uk,Guth:1982pn,Enqvist:1991xw,Turner:1992tz}
	\begin{equation}
	P = e^{-I}.
	\end{equation}	
The change in the physical false vacuum volume, $\mathcal{V}_{\rm false} = a^{3} P(T)$, normalized to the Hubble rate, is captured by the equation~\cite{Turner:1992tz} 
	\begin{equation}
	\frac{1}{H \mathcal{V}_{\rm false}} \frac{d\mathcal{V}_{\rm false}} {dt} = 3 + T \frac{dI} {dT}.
	\end{equation} 
We define the percolation temperature $T_p$ as the highest temperature for which both $I(T) > 1$ and $d\log\mathcal{V}_{\rm false}/dt < -H$ hold (sometimes slightly less stringent conditions are employed, e.g.~see discussion in~\cite{Ellis:2018mja}, the precise choice does not affect our qualitative results below). Numerically, we find the first condition, $I(T) > 1$, occurs later and hence determines the percolation condition in our parameter space.  

To characterise the strength of the phase transition, we find the free energy difference between the true and false vacuua normalized to the radiation density,
	\begin{equation}
	\alpha = \frac{1}{\rho_{\rm rad}}\left( 1 - T \frac{\partial }{\partial T} \right) \bigg( V_{\rm eff}( 0, T) - V_{\rm eff}( \phi_{n}, T)  \bigg)\bigg|_{T=T_{\ast}},
	\end{equation}
where $T_{\ast}$ is set to either the nucleation or percolation temperature.\footnote{ When eventually calculating the entropy dilution factor below, evaluating $\alpha$ at $T_{\ast}=T_p$ gives a better estimate, due to the additional expansion between $T_n$ and $T_p$. The choice is irrelevant when determining the GW signal for the PTs of interest as $\alpha \gg 1$. On the other hand, when approximating the mean bubble radius using $\beta_{H}$, we numerically find evaluating $\beta_{H}$ at $T_{\ast}=T_n$ gives a more accurate result, after comparing to the mean radius extracted from the bubble distribution itself. } The inverse timescale of the phase transition normalized to the Hubble rate reads
	\begin{equation}
	\beta_{H} \equiv \frac{ \beta }{H }\bigg|_{t=t_{\ast}} \equiv \frac{1}{H \Gamma_{\rm bub}}\frac{d\Gamma_{\rm bub}}{dt}\bigg|_{t=t_{\ast}}  = - \frac{T}{ \Gamma_{\rm bub}}\frac{d\Gamma_{\rm bub}}{dT}\bigg|_{T=T_{\ast}}.
	\end{equation}
By numerically calculating the bubble action $S$ for a sufficiently large number of points over a suitable temperature range, we obtain a good approximation of $\Gamma_{\rm bub}(T)$, which allows us to find all the above quantities. Phase transition parameters as a function of $g_{\BL}$ are shown in Fig.~\ref{fig:PTexample}. Clearly in the limit of a small $\beta_{\lambda_\rho}$ we have strong supercooling, with $\beta_{H} \lesssim 8$ and $\alpha \gg 1$, which allows for cosmologically significant PBH production, e.g.~as found in~\cite{Gouttenoire:2023naa}. We now turn to calculating the PBH abundance.

\section{Primordial black hole production}

\subsection{Background evolution}

We first consider an average Hubble patch. The total energy density, $\rho,$ consist of the vacuum energy density, $\rho_{\rm vac}$, and the radiation, $\rho_{\rm rad}$. The latter is made up of the plasma of relativistic particles together with the bubble walls moving at $v_{w} \simeq 1$~\cite{Gouttenoire:2023naa}. The Friedmann equations are given by
	\begin{subequations}
	\begin{align}
	H_{\rm bkg}^{2} & = \frac{8 \pi}{3} \frac{\rho_{\rm vac} + \rho_{\rm rad} }{M_{\Pl}^2}, \\
	\frac{d \rho_{\rm rad} }{dt } & = -4H_{\rm bkg} \rho_{\rm rad} - \frac{ d \rho_{\rm vac} }{ dt},
	\end{align}
	\end{subequations}
where $H_{\rm bkg}$ denotes the Hubble rate of the average background patch.
We can relate $\rho_{\rm vac}$ to the bubble nucleation rate through
	\begin{equation}
	\rho_{\rm vac} = \Lambda_{\rm vac} e^{-I}.
	\end{equation}
To solve the Friedmann equations, we find it useful to switch coordinates, from time to false vacuum temperature. We then have 
	\begin{subequations}
	\begin{align}
	H_{\rm bkg}^{2} & = \frac{8 \pi}{3} \frac{\rho_{\rm vac} + \rho_{\rm rad} }{M_{\Pl}^2}, \\
	\frac{ d \rho_{\rm rad} }{ dT } & = \frac{ 4 H_{\rm bkg} \rho_{\rm rad} }{HT} - \frac{d \rho_{\rm vac} }{dT} ,
	\end{align}
	\end{subequations}
where $T$ is the temperature in the false vacuum, and which we solve to find $\rho_{\rm rad}(T)$. Note $\rho_{\rm vac}$, through the quantity $I$ which contains the bubble radius, effectively also contains the scale factor. Hence the equations should be solved self-consistently by evaluating $I$ by taking into account the scale factor at each time step, which in turn can be found through $H_{\rm bkg}$. In order to solve the equations, however, a simplification is possible. First note that before percolation, $H_{\rm bkg}$ is for practical purposes given by its corresponding false vacuum value. After percolation, the $\rho_{\rm vac}$ source term for the radiation is, by definition, approximately negligible. We therefore first evaluate $I$ numerically using the scale factors and Hubble rate in the false vacuum, $H$, and use this to solve the Friedmann equations including $H_{\rm bkg}$ of the background patch in the first equation. We then check whether the approximation is a good one by using the resulting scale factor in an updated determination of $I$. The resulting shift in $T_{p}$ is tiny --- we will quantify it below --- and so we confirm the approximation is justified.

For our calculations, we will also need the scale factor as a function of the false vacuum temperature. Ignoring changes in degrees-of-freedom, for the scale factor in a false vacuum patch we have the standard relation,
	\begin{equation}
	\frac{ a(T) }{ a(T') } = \frac{ T' }{ T }.
	\label{eq:falsevacscale}
	\end{equation}
For the scale factor in the background patch that has nucleated, we instead solve 
	\begin{equation}
	\frac{ d a_{\rm bkg} }{d T } = - \frac{ a_{\rm bkg} }{ T } \frac{ H_{\rm bkg} }{ H }.
	\end{equation}
The equation is solved starting from a sufficiently high temperature where $H_{\rm bkg} = H$ and $a_{\rm bkg}(T)=a(T)$. Note, for bookkeeping purposes, the temperatures which appear here should be understood as the temperature in the false vacuum patch, not the temperature in the nucleating background patch, which contains nucleated bubbles and partially reheated plasma. Using our solution, it is convenient to define
	\begin{equation}
	A(T,T') \equiv \frac{ a_{\rm bkg}(T) T }{ a_{\rm bkg}(T') T' } , 
	\end{equation}
which changes from unity to take into account deviations from the false vacuum relation, Eq.~\eqref{eq:falsevacscale}, for the background scale factor.

In terms of the false vacuum temperatures and false vacuum H, the bubble radius in the background patch is then more precisely given by
	\begin{equation}
	R_{\rm bkg}(T,T') = \frac{1}{T}\int_{T}^{T'} \frac{ A(T,\tilde{T}) d\tilde{T} }{ H(\tilde{T}) }.
	\label{eq:bkgbubblesize}
	\end{equation}
It can also be of interest to consider the bubble density as a function of radius~\cite{Turner:1992tz,Ellis:2018mja}, which rewritten as a function of the false vacuum temperature is given by 
	\begin{equation}
	\frac{dn_{\rm bkg}}{dR_{\rm bkg}}(T,R_{\rm bkg}) = \frac{dT'}{dR_{\rm bkg}}\frac{T^{3} \Gamma_{\rm bub}(T')P(T')}{  A(T,T')^3 T'^{4} H(T') } .
	\end{equation}
where here $T' \equiv T'(R)$ is understood as being the false vacuum temperature at which a bubble of physical size $R_{\rm bkg}$ was nucleated, found by numerically inverting Eq.~\eqref{eq:bkgbubblesize}, from which one also finds the corresponding derivative.

\subsection{Late patch evolution}

The above describes the evolution of an average background nucleating patch. We are now interested in the evolution of a late nucleating patch. We assume no bubble nucleates in the patch until some temperature $T_{i} < T_{n}$.  By modifying the high temperature terminal in Eq.~\eqref{eq:Ifunc}, we find the vacuum energy in the late patch,
	\begin{align}
	I_{\rm late} = \frac{4\pi}{3}\int_{T}^{T_{i}}dT'\frac{ \Gamma_{\rm bub}(T') }{ T'^{4}H(T') }\left(\int_{T}^{T'} \frac{ d\tilde{T} }{ H(\tilde{T}) } \right)^{3}.
	\label{eq:Ifunclate}
	\end{align}
Again we use the same approximation as before, in assuming $H=H_{\rm false}$, when calculating $I_{\rm late}$. The probability of finding a point in the false vacuum in the late patch is then
	\begin{equation}
	P_{\rm late} = e^{-I_{\rm late}}.
	\end{equation} 
Thus the vacuum energy density in the late patch is simply
	\begin{equation}
	\rho_{\rm vac late} = \Lambda_{\rm vac} e^{-I_{\rm late}}.
	\end{equation}
We then also solve the Friedmann equations for the late patch 
	\begin{subequations}
	\begin{align}
	H_{\rm late}^{2} & = \frac{8 \pi}{3} \frac{\rho_{\rm vac late} + \rho_{\rm rad late} }{M_{\Pl}^2}, \\
	\frac{ d \rho_{\rm rad late} }{ dT } & = \frac{ 4 H_{\rm late} \rho_{\rm rad late} }{HT} - \frac{d \rho_{\rm vac late} }{dT} .
	\end{align}
	\end{subequations}
Eventually we will also need the scale factor in the late patch, found by solving
	\begin{equation}
	\frac{ d a_{\rm late} }{d T } = - \frac{ a_{\rm late} }{ T } \frac{ H_{\rm late} }{ H }.
	\end{equation}
Similarly to what we did for the background patch, it is convenient to define
		\begin{equation}
	B(T,T') \equiv \frac{ a_{\rm late}(T) T }{ a_{\rm late}(T') T' } , 
	\end{equation}
to parametrize deviations from the false vacuum scale factor/temperature relation. The bubble size in the late patch is then given by 
	\begin{equation}
	R_{\rm late}(T,T') = \frac{1}{T}\int_{T}^{T'} \frac{ B(T,\tilde{T}) d\tilde{T} }{ H(\tilde{T}) }.
	\label{eq:latebubblesize}
	\end{equation}
The bubble spectrum of the late patch is given by 
	\begin{equation}
	\frac{dn_{\rm late}}{dR_{\rm late}}(T,R_{\rm late}) = \frac{dT'}{dR_{\rm late}}\frac{T^{3} \Gamma_{\rm bub}(T')P_{\rm late}(T')}{  B(T,T')^3 T'^{4} H(T') } \Theta(T_i - T') .
	\end{equation}
where the $\Theta$ function takes into account that no bubbles formed at $T' > T_{i}$ in the late patch.

\subsection{Collapse and fractional abundance}

For each choice of $T_{i}$ we can also calculate the contrast density in radiation
	\begin{equation}
	\delta(T) \equiv \frac{ \rho_{\rm rad late} - \rho_{\rm rad } }{ \rho_{\rm rad } }.
	\end{equation}
This reaches a maximum shortly after late patch percolation, as the energy density in the background patch has began to become diluted a little earlier, while the late patch energy density is still constant due to the vacuum. After late patch percolation, $\delta$ decreases again as $H_{\rm late} > H_{\rm bkg}$ leading to a faster redshifting. We define $T_{\delta \mathrm{max}}$ as the temperature at which $\delta$ is maximized, $\delta_{\rm max} \equiv \delta(T_{\delta \mathrm{max}})$. The smaller $T_{i}$ the larger $\delta_{\rm max}$. 

We assume a patch collapses if $\delta_{\rm max}$ reaches a threshold value, $\delta_{c}$, for critical collapse. Calculations, some based on full general relativistic simulations, indicate $0.4 \leq \delta_{c} \leq 0.66$ in the context of overdensities from inflation re-entering the Hubble horizon~\cite{Carr:1975qj,Shibata:1999zs,Musco:2004ak,Harada:2013epa,Musco:2018rwt,Musco:2020jjb,Germani:2018jgr,Musco:2018rwt,Escriva:2019phb,Escriva:2021pmf,Escriva:2022bwe,Stamou:2023vxu}. The precise value of $\delta_{c}$ depends on the shape of the overdensity, but its effects can be captured by considering the maximum of the so-called compaction function and its curvature, and thus be related to the power spectrum of curvature perturbations. Note the calculations have been performed assuming spherical symmetry and typically also isotropic pressure (although see~\cite{Musco:2021sva}). We expect departures from these assumptions in the PT scenario, due to the nature of the Hubble patch just after bubble percolation. Note non-sphericity is expected to increase $\delta_{c}$~\cite{Sheth:1999su,Baumgarte:2015aza,Clough:2016jmh,Kuhnel:2016exn,Sasaki:2018dmp}. The value of $\delta_{c}$ will have an effect on which input parameters return $f_{\PBH} =1$, but the strong GW signal will not be sensitive to our precise choice. In this work, we follow previous PT literature~\cite{Gouttenoire:2023naa}, and take $\delta_{c} = 0.45$ to aide comparison. The late patch nucleation temperature, $T_i$, is thus from now on fixed by requiring the maximal density contrast, $\delta_{\rm max}$,  be equal to the critical collapse threshold, $\delta_{c}$.

We assume the collapse occurs at $T_{\delta \mathrm{max}}$ with corresponding time $t_{\delta \mathrm{max}}$. The probability of the horizon size patch, which eventually collapses into a PBH, having no bubbles at $T_{i}$ is given by~\cite{Liu:2021svg}
	\begin{subequations}
	\begin{align}
	P_{\rm no \; bub} &  = \mathrm{Exp}\left[ - \int_{t_c}^{t_i} dt' \Gamma_{\rm bub}(t') a_{\rm late}(t')^{3} V_{\rm no \; bub}  \right] \\
		          &  = \mathrm{Exp}\left[ - \int_{T_i}^{T_c} \frac{ dT' \Gamma_{\rm bub}(T') }{ T' H(T') } a_{\rm late}(T')^{3} V_{\rm no \; bub} \right],
	\end{align} 
	\end{subequations}
where the temperature and Hubble rate are those of the false vacuum, and the volume factor,
	\begin{equation}
	V_{\rm no \; bub} = \frac{ 4\pi }{3 } \left[ \frac{1}{ a_{\rm late}(t_{\delta \mathrm{max}})H_{\rm late}(t_{\delta \mathrm{max}})} \right]^3 ,
	\end{equation}
represents the comoving volume of the Hubble sized patch at the start of the collapse.
However, the calculation of $\delta$ for the critical collapse also assumes no larger bubbles --- from background patches surrounding the late nucleating patch --- enter into the collapsing volume before $\delta_{\rm max}$ is attained. Thus, it has been advocated~\cite{Gouttenoire:2023naa}, that the collapse probability is better estimated as
	\begin{subequations}
	\begin{align}
	P_{\rm coll} & =  \mathrm{Exp}\left[ - \int_{t_c}^{t_i} dt' \Gamma_{\rm bub}(t') a_{\rm late}(t')^{3} V_{\rm coll} \right] \\
		     & = \mathrm{Exp}\left[ - \int_{T_i}^{T_c} \frac{ dT' \Gamma_{\rm bub}(T') }{ T' H(T') } a_{\rm late}(T')^{3} V_{\rm coll} \right], \label{eq:Pcoll}
	\end{align}
	\end{subequations}
where the volume factor is
	\begin{subequations}
	\begin{align}
	V_{\rm coll} & = \frac{ 4\pi }{ 3 } \left[ \frac{ 1 } { a_{\rm late}(t_{\delta \mathrm{max}})H_{\rm late}(t_{\delta \mathrm{max}})} + r(t_{\delta \mathrm{max}},t')  \right]^{3} \\
		     & = \frac{ 4\pi }{ 3 } \left[  \frac{ 1 } { a_{\rm late}(T_{\delta \mathrm{max}})H_{\rm late}(T_{\delta \mathrm{max}})} + \int_{T_{\delta \mathrm{max}}}^{T'} \frac{ d\tilde{T} }{ \tilde{T}H(\tilde{T})a_{\rm bkg}(\tilde{T}) }  \right]^{3}. \label{eq:volfactor}
	\end{align}
	\end{subequations}
Here, in the evaluation of $r(t_{\delta \mathrm{max}},t')$, Eq.~\eqref{eq:comovingrad}, we use the background value for the scale factor.
The eventual PBH mass is estimated as the energy inside the sound horizon of the collapsing patch at $t_{\delta \mathrm{max}}$ (e.g.~see~\cite{Fujita:2014hha})
	\begin{equation}
	M_{\PBH} = \frac{4 \pi (\rho_{\rm rad late}+\rho_{\rm vac late}) c_{s}^{3}}{3 H_{\rm late}^{3}} = \frac{c_{s}^{3}}{2} \frac{ M_{\rm Pl}^2}{ H_{\rm late} },
	\end{equation}
where $c_{s} = 1/\sqrt{3}$ is the sound speed (although included, $\rho_{\rm vac late}$ is negligible at $t_{\delta \mathrm{max}}$). The current study is limited to the above monochromatic estimate, we leave for future work the full derivation of a detailed mass spectrum taking into account the full critical collapse phenomenon~\cite{Choptuik:1992jv,Niemeyer:1997mt,Green:1999xm,Musco:2008hv,Kuhnel:2015vtw}. Some preliminary estimates, showing the monochromatic estimate is a good one, are given in App.~\ref{sec:nonmono}.

\begin{figure}[p]
\begin{center} 
\includegraphics[width=210pt]{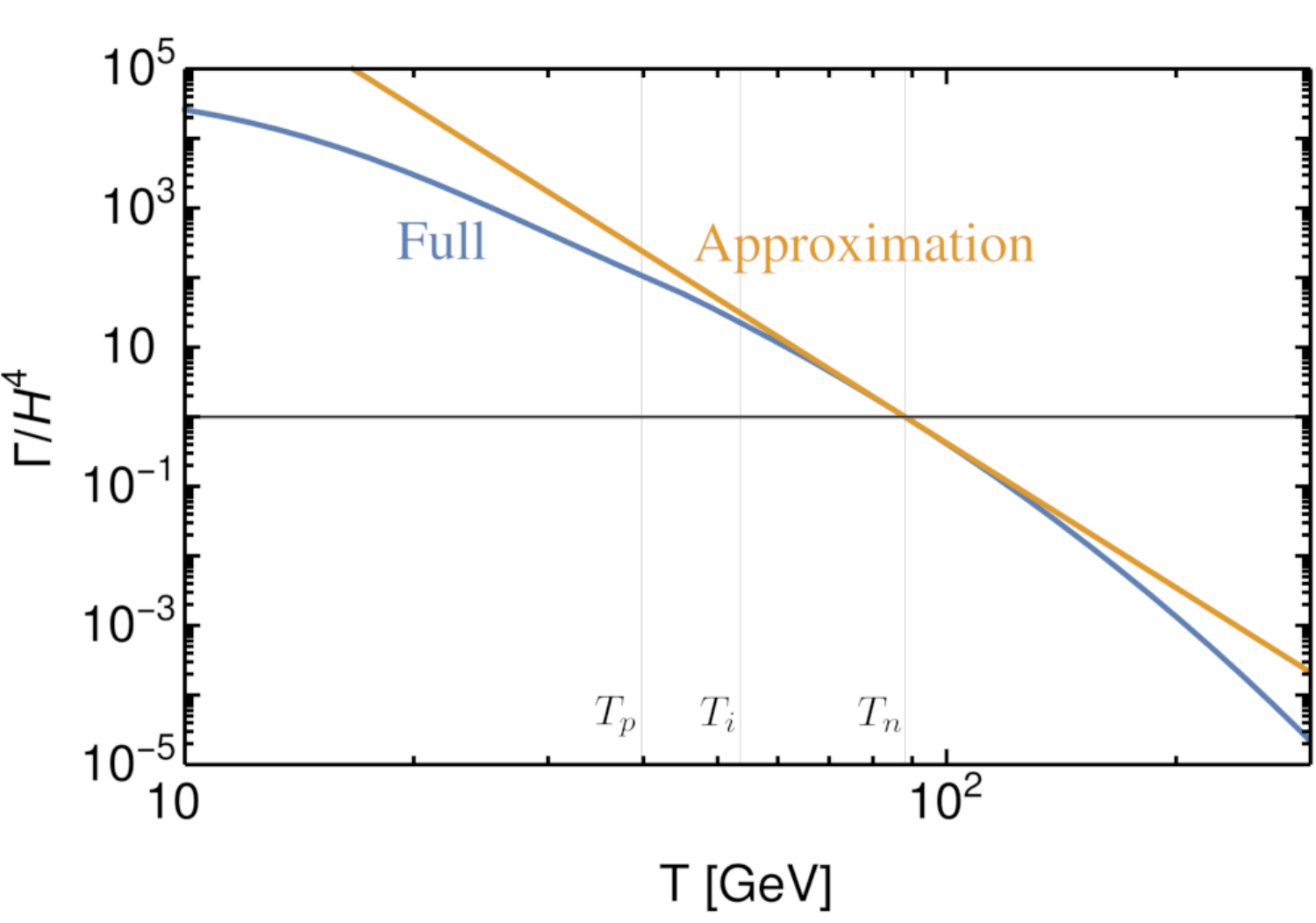} $\quad$
\includegraphics[width=210pt]{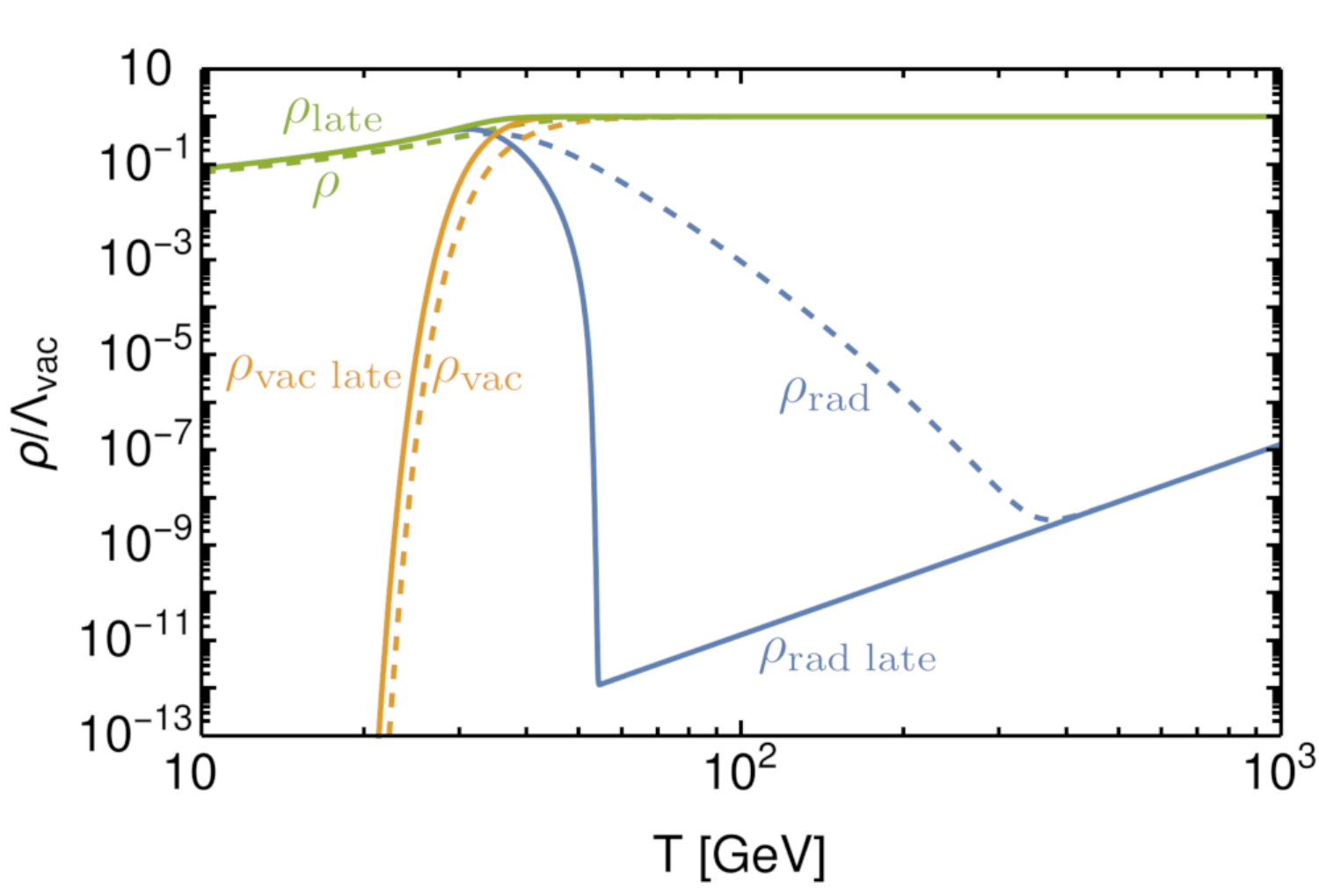}
\\
\includegraphics[width=210pt]{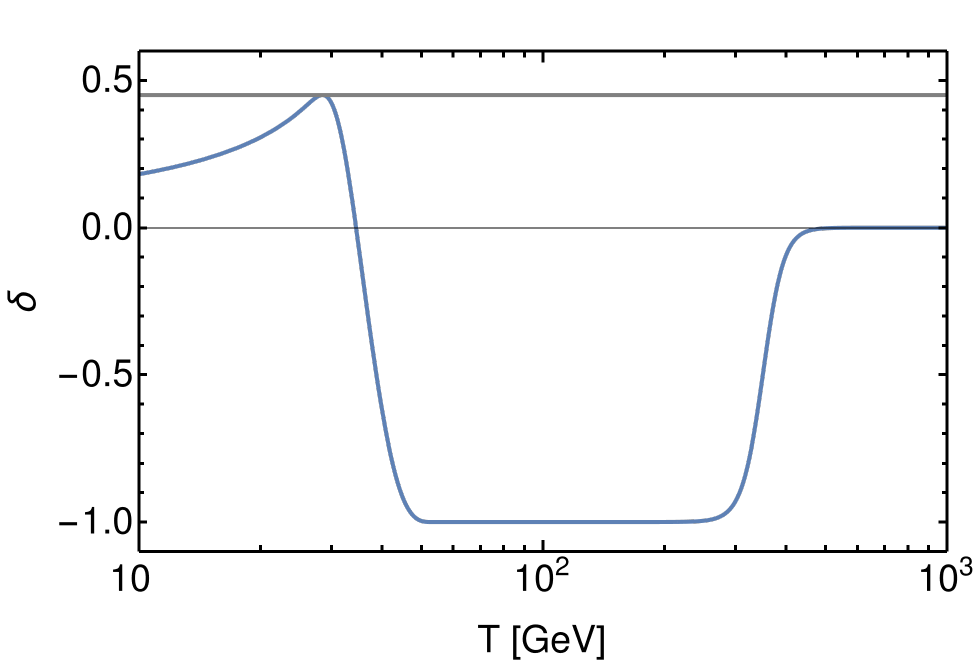} $\quad$
\includegraphics[width=210pt]{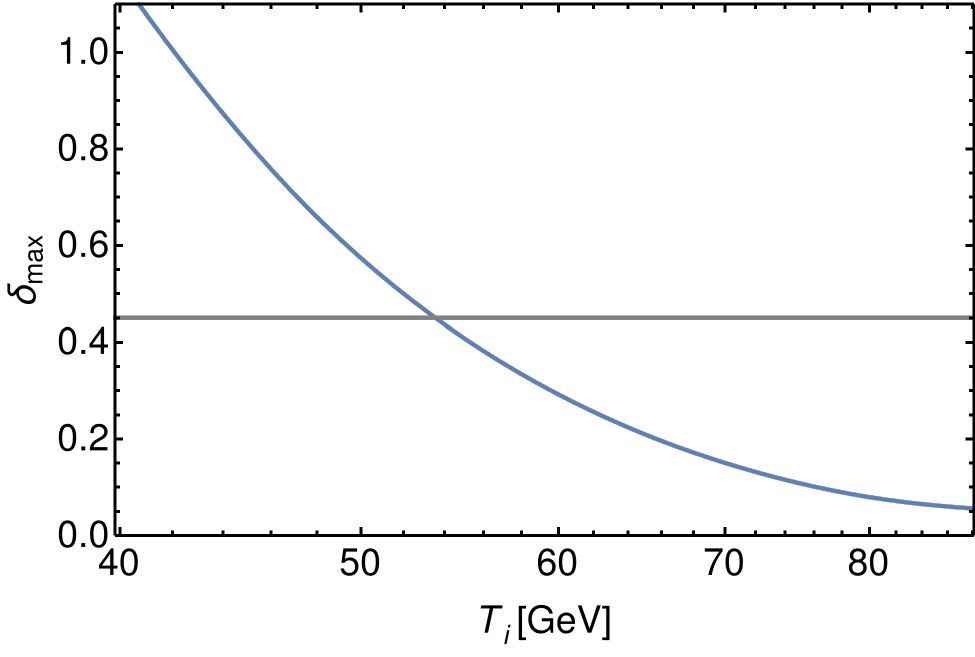} 
 \\
\includegraphics[width=210pt]{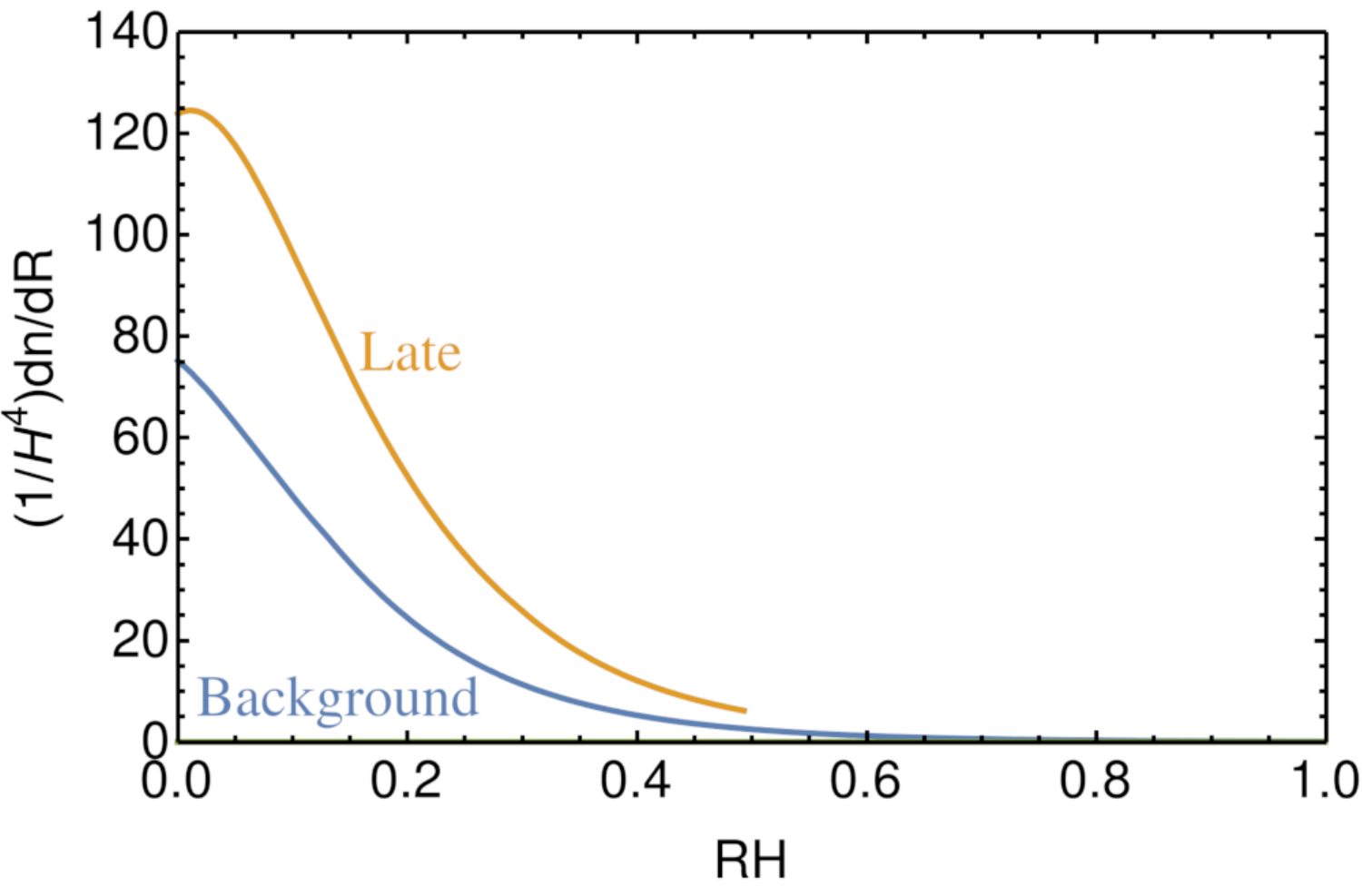}  $\quad$
\includegraphics[width=210pt]{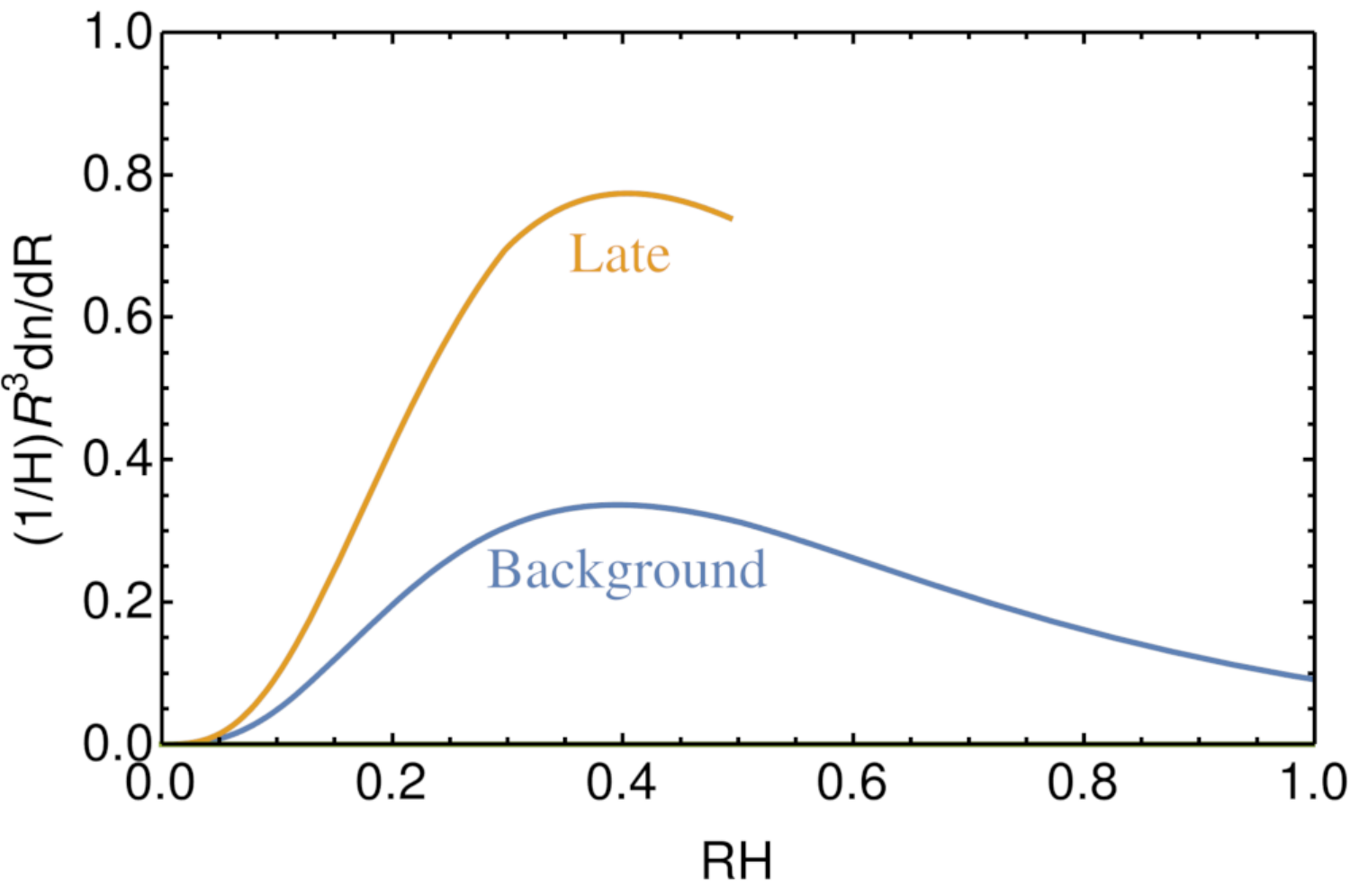} 
\end{center}
\caption{\small Characterisation of several important quantities for PBH formation and the phase transition dynamics for a bechmark scenario with $v_{\rho} = 10^{3}$ TeV, $y_{Ni} = 0.01$, and $g_{\BL} = 0.2958 $ that predict $f_{\PBH} =1$, with $M_{\PBH} \simeq 4.5 \times 10^{-12} \; M_{\odot}$. We depict the behaviour of the nucleation rate comparing the full calculation of the nucleation rate with the approximation often used in the literature~\cite{Liu:2021svg,Gouttenoire:2023naa} (top left), energy densities (top right) and density contrast (middle left, with $T_{i} \simeq 53.6 \; \mathrm{GeV}$) as a function of the false vacuum temperature $T$. We also show the density contrast as a function of $T_i$ (middle right). Finally we display the bubble distribution at percolation (bottom left), where $RH$ is the bubble size as a fraction of the Hubble length. Weighting the distribution by $R^{3}$ we see the differences in the large bubbles between the background and late patches (bottom right).  }
\label{fig:densities}
\end{figure}

We now wish to find, $f_{\PBH}$, the PBH-to-DM fraction at late times. To do this we note that PBH formation occurs a little after bubble percolation, around time $t_{\delta \rm max}$ when the false vacuum temperature is $T_{\delta \rm max}$, and we denote the corresponding background temperature as $T_{\rm bkg \, form}$ (note $T_{\rm bkg \, form} \approx T_{\RH} \approx T_{\rm infl}$ with small differences found numerically and taken into account in our results). The ratio of densities at this time is
	\begin{equation}
	\frac{ \rho_{\PBH}(t_{\delta \rm max}) }{ \rho_{\rm rad}(t_{\delta \rm max}) } \simeq c_{s}^{3} P_{\rm coll} \frac{ H_{\rm bkg}(t_{\delta \rm max}) }{ H_{\rm late}(t_{\delta \rm max}) } ,
	\end{equation}
where the ratio of Hubble rates is a correction which takes into account that $M_{\PBH}$ is set by the late rather than background patch density.
Later, at matter radiation equality, $T_{\rm eq} \simeq 0.8$ eV, we have $\rho_{\rm rad} = \rho_{\rm m} = \rho_{\DM} (\Omega_{\DM}+\Omega_{B})/\Omega_{\DM}  \simeq 1.2 \rho_{\DM}$. Between PBH formation, at $T \simeq T_{\rm bkg \, form}$, and $T_{\rm eq}$, one has $\propto 1/a^{3}$ dilution so
	\begin{equation}
	\frac{ \rho_{\PBH}(T_{\rm eq})}{ \rho_{\PBH}(T_{\rm bkg \, form}) } =  \left(\frac{ a_{\rm bkg}(T_{\rm bkg \, form}) }{  a_{\rm bkg}(T_{\rm eq}) }\right)^{3}.
	\end{equation}
From entropy conservation we have
	\begin{equation}
	\frac{ a_{\rm bkg}(T_{\rm bkg \, form}) }{ a_{\rm bkg}(T_{\rm eq}) } = \left( \frac{ g_{\ast s}(T_{\rm eq}) }{ g_{\ast s}(T_{\rm bkg \, form}) } \right)^{1/3} \frac{ T_{\rm eq} }{ T_{\rm bkg \, form} },
	\end{equation}
where $g_{\ast s}$ are the entropic degrees-of-freedom.
The radiation instead redshifts as $\propto 1/a^{4}$, but it also receives reheating contributions, so
	\begin{equation}
	\frac{ \rho_{\rm rad}(T_{\rm eq})}{ \rho_{\rm rad}(T_{\rm bkg \, form}) } = \frac{ g_{\ast}(T_{\rm eq})T_{\rm eq}^{4} } { g_{\ast}(T_{\rm bkg \, form})T_{\rm bkg \, form}^{4}}
	\end{equation}
Combing all the above we have
	\begin{equation}
	f_{\PBH} = \frac{ \rho_{\PBH} }{ \rho_{\DM} } \simeq  c_{s}^{3} \, P_{\rm coll} \, \frac{\Omega_{\DM}+\Omega_{B}}{ \Omega_{\DM}} \, \frac{ g_{\ast s}(T_{\rm eq}) }{ g_{\ast }(T_{\rm eq})  } \frac{ T_{\rm bkg \, form} }{ T_{\rm eq} } \frac{ H_{\rm bkg}(t_{\delta \rm max}) }{ H_{\rm late}(t_{\delta \rm max}) } ,
	\end{equation}
where we have used $g_{\ast}(T_{\rm bkg \, form}) = g_{\ast s}(T_{\rm bkg \, form})$, and we remind the reader that $g_{\ast s}(T_{\rm eq}) = 3.91$ and $g_{\ast }(T_{\rm eq}) = 3.36$.

\subsection{Application to example model}

Having developed the above machinery, we now calculate the PBH fraction for our example $B-L$ model, and search for parameter space in which $f_{\PBH}=1$. In Fig.~\ref{fig:densities}, we characterise PBH formation and the strongly first-order phase transition for a benchmark point for which $v_{\rho} = 10^{3}$ TeV, $y_{Ni} = 0.01$, and $g_{\BL} = 0.2958 $. These values were judiciously chosen so as to return $f_{\PBH} =1$. The PBH mass is $M_{\PBH} \simeq 4.5 \times 10^{-12} \; M_{\odot}$. In the top left plot we show our full determination of $\Gamma_{\rm bub}(T)$ and compare it with the approximate form, Eq.~\eqref{eq:gammaapprox}, rewritten in terms of temperature 
	\begin{equation}
	\Gamma_{\rm bub}(T) \approx H(T_n)^{4}\left(\frac{ T_n }{ T } \right)^{\beta_{H}(T_n)}.
	\end{equation}
Deviations between the full and approximate expression are visibly present at $T = T_{i} \approx 54 \; \mathrm{GeV}$ which returns the required contrast for PBH formation. We will eventually show the resulting modest differences in the $\beta_{H}$ required to achieve $f_{\PBH}=1$ below.

In Fig.~\ref{fig:densities} top right we first show the evolution of energy densities with the false vacuum temperature $T$. The dashed lines in the plot represent the vacuum, radiation, and total energy densities of an average Hubble patch, while the solid lines illustrate these same quantities for a late-nucleating patch. At sufficiently high false vacuum temperatures, the radiation energy densities for both types of Hubble patches align, and the same is observed with the vacuum energy densities. As nucleation starts within an average Hubble patch, its radiation energy density begins to increase relative to that of the late-nucleating patch. Once the percolation threshold of the true vacuum phase is achieved, there is a drastic drop in the vacuum energy density, leading the average background patch to become dominated by radiation. Concurrently, the energy density within this patch begins to dilute due to expansion, whilst the late-nucleating patch remains vacuum dominated. Once the percolation threshold is reached within the late-nucleating patch, an excess of radiation energy density relative to the surrounding background Hubble patches follows, triggering the collapse into a PBH.

In the middle row of Fig.~\ref{fig:densities}, we show the dependence of $\delta_{\rm max}$ on $T_{i}$ (right) and the dependence of the density contrast with the false vacuum temperature $T$ (left). We observe how we start off from a homogeneous state at sufficiently high false vacuum temperatures. The density contrast drops to its minimum value approximately when bubbles start nucleating in the background Hubble patches and increases again when nucleation starts in the delayed patch. Ultimately, it reaches a maximum shortly after late patch percolation, as the energy density in the
background patch has began to become diluted a little earlier, while the late patch energy
density is still constant due to the vacuum.

Finally, in the bottom row  we show the bubble distribution of the average background patch and the late patch with $\delta_{\rm max} = 0.45$ at their respective percolation temperatures. The late patch features approximately twice the number of bubbles as the background patch. Due to the first bubbles being delayed, the distribution of bubbles in the late patch features a cut in the distribution, here at roughly half the Hubble length. In the right plot, we weight the distribution by $R^3$ to show the volume occupied by the bubble. Therein we see a significant volume is occupied by rare large bubbles, especially for the background patch, which explains why the late patch is filled instead by roughly twice as many small bubbles.

\begin{figure}[p]
\begin{center} 
\includegraphics[width=200pt]{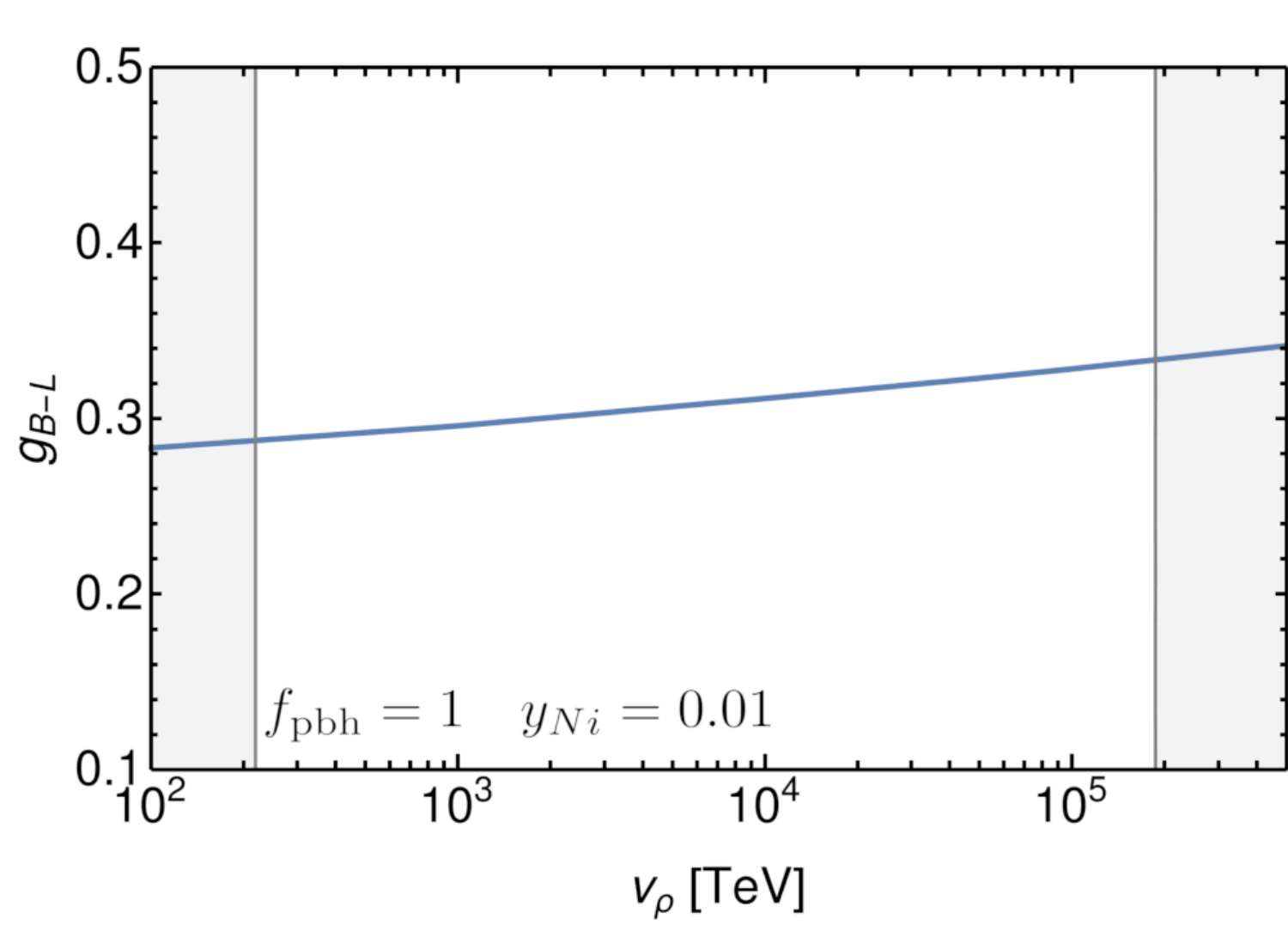} $\quad$
\includegraphics[width=210pt]{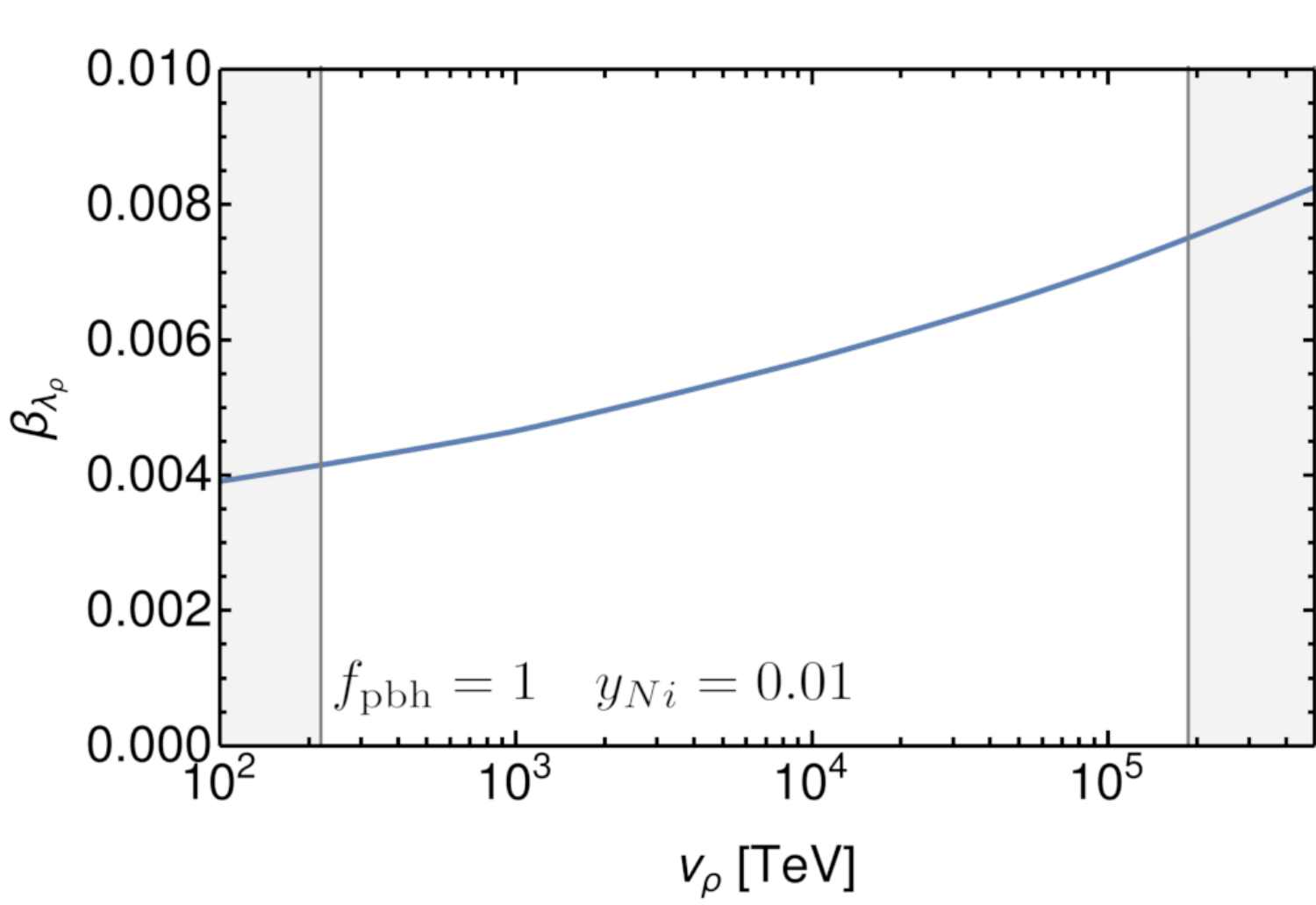} \\
\includegraphics[width=210pt]{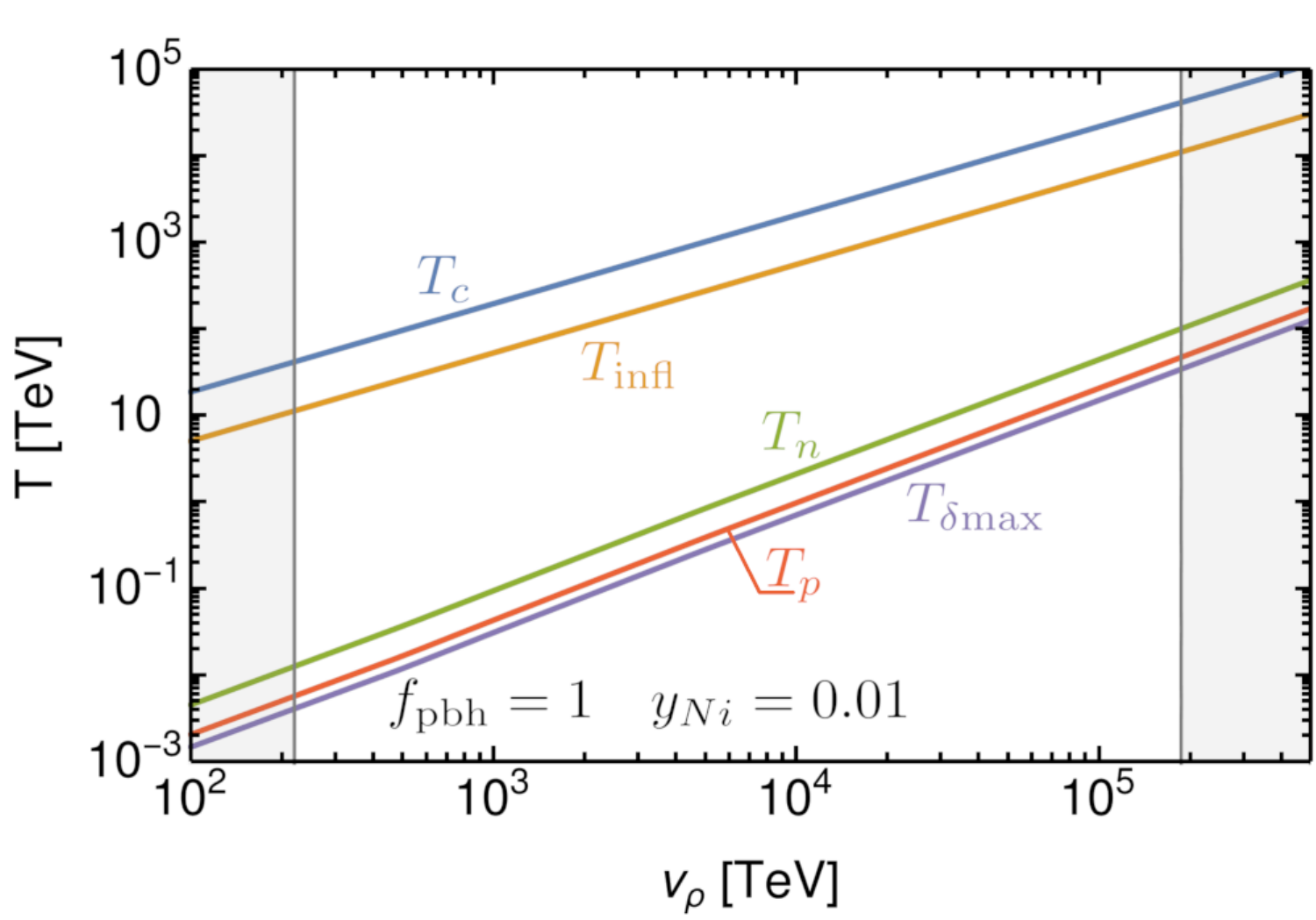}
\includegraphics[width=210pt]{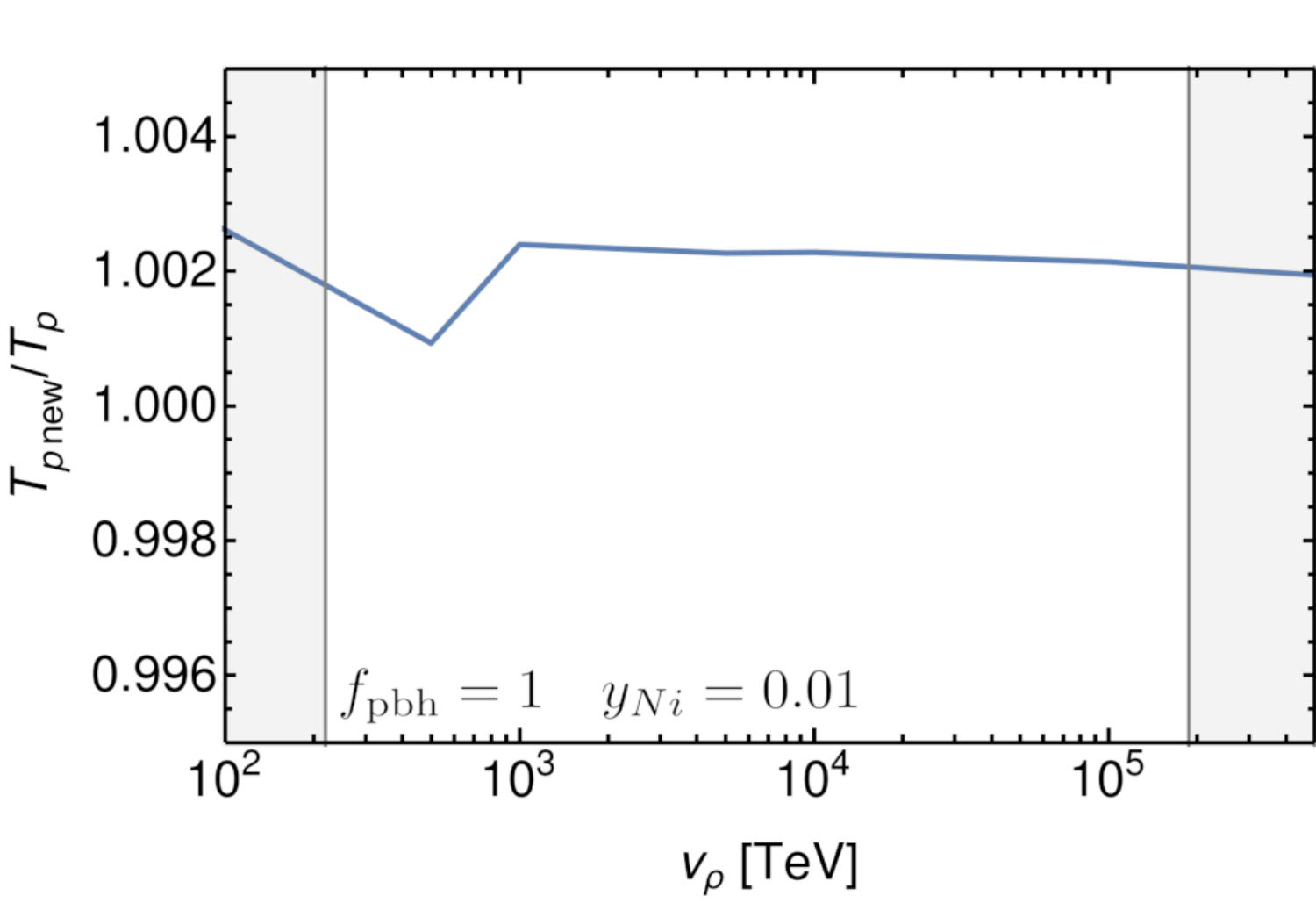} \\
\includegraphics[width=210pt]{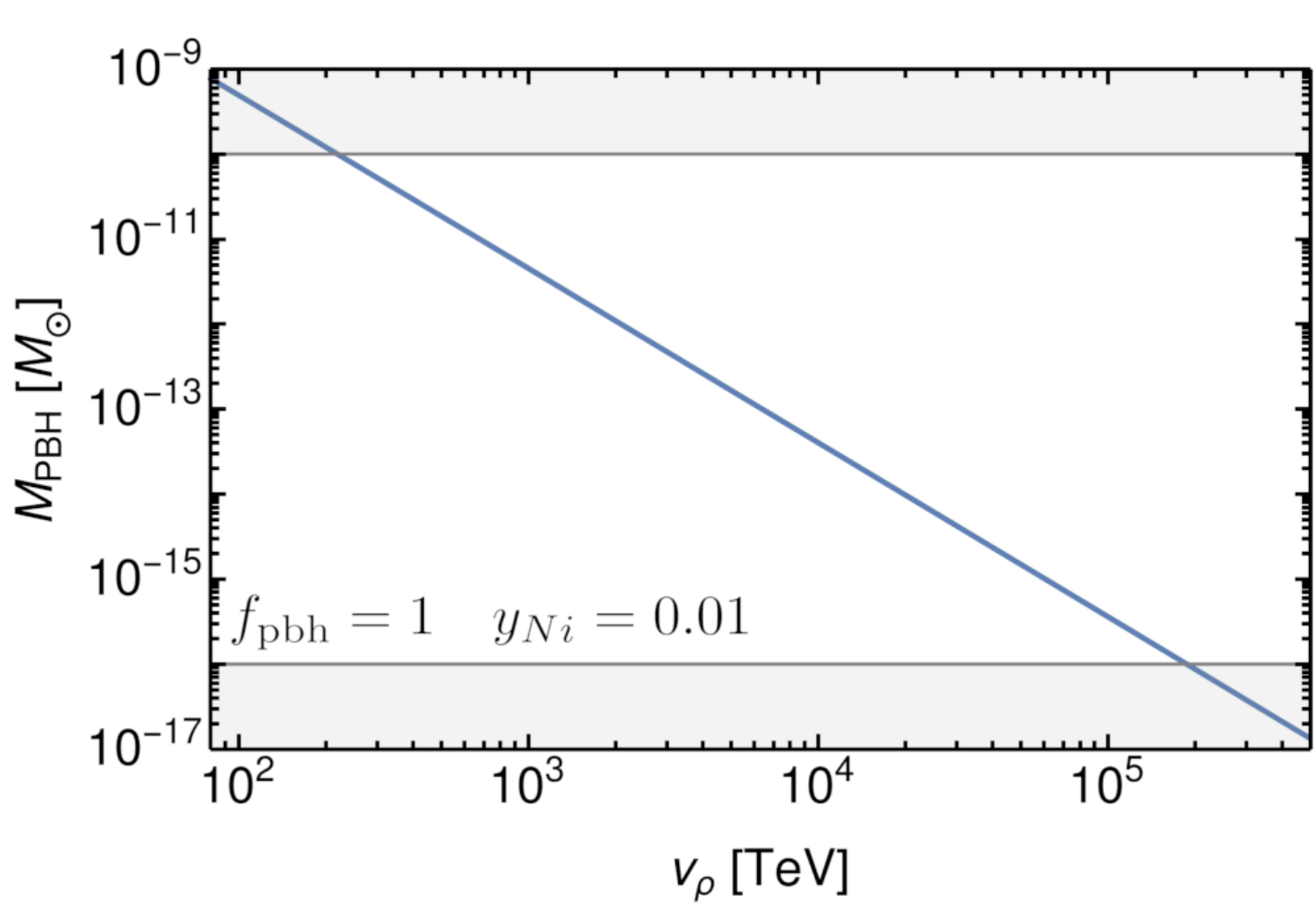}
\includegraphics[width=210pt]{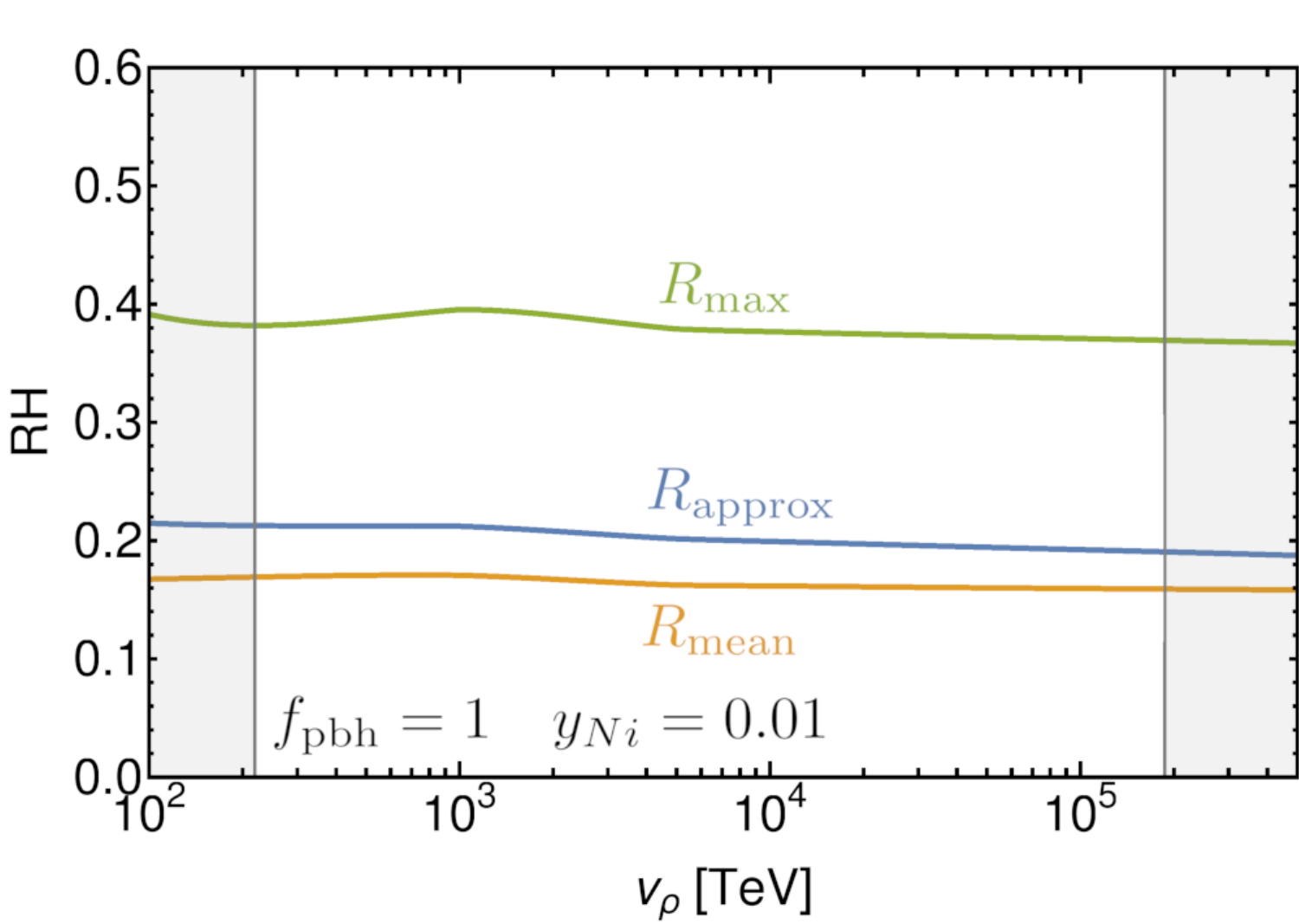}
\end{center}
\caption{\small Top: The required gauge coupling (left) and associated beta function (right) giving $f_{\PBH}=1$ as a function of the $B-L$ vev, $v_{\rho}$. Shaded regions are excluded by PBH constraints. Middle: Various temperatures characterizing the PT as a function of $v_{\rho}$. Note the new percolation temperature $T_{p \mathrm{new}}$ following our iterative check shows only a per mille correction with respect to $T_{p}$ (right). Bottom: The resulting PBH mass (left) and the bubble radii at collision (right), namely the mean radius $R_{\rm mean}$, the approximate radius, $R_{\rm approx} = \pi^{1/3}/\beta(T_n)$~\cite{Enqvist:1991xw}, and the radius at which the bubble energy (and volume) distribution is maximized, $R_{\rm max}$~\cite{Ellis:2018mja}.}
\label{fig:gblreq}
\end{figure}

Now we turn to Fig.~\ref{fig:gblreq}. For a given choice of $v_{\rho}$, we can scan over $g_{\BL}$ to find where $f_{\PBH}=1$ (we always set $y_{i} = 0.01$ as an example). The required $g_{\BL}$ to achieve $f_{\PBH} =1$ is shown in Fig.~\ref{fig:gblreq} (top left) together with the corresponding value of the $\beta_{\lambda_{\rho}}$ function (top right).

In the middle panel of Fig.~\ref{fig:gblreq} we show various temperatures characterizing the PT. The percolation temperature and also $T_{\delta \mathrm{max}}$ remains above QCD confinement, so we do not need to take into account QCD enhancement of $\Gamma_{\rm bub}$, present at lower temperatures~\cite{Witten:1980ez,Iso:2017uuu,vonHarling:2017yew,Sagunski:2023ynd}. A smaller $\beta_{\lambda_\rho}$ corresponds to a smaller vacuum energy difference between the two phases, which strengthens the phase transition~\cite{Dorsch:2017nza}. Furthermore, smaller values of $g_{\BL}$ imply a weaker running of couplings which, in turn, delays the transition time and makes the derivative of the bubble action $S$ smaller~\cite{Jinno:2016knw}.

As mentioned above, before solving the Friedmann equations, we first fix $H = H_{\rm false}$ in our calculation of $I(T)$. To check this is a good approximation, we then iterate, using our $H_{\rm bkg}$ determination to calculate a new $I(T)$, and use this to find an updated percolation temperature $T_{p \, \mathrm{new}}$. The ratio of the two is displayed in Fig.~\ref{fig:gblreq} middle right, showing only very small changes in $T_{p}$, thus justifying the approximation. In Fig.~\ref{fig:gblreq} bottom we shown the resulting $M_{\PBH}$ and various measures of the bubble radius at collision.

In the bottom row on the left, we show how the PBH mass decreases as we go to larger vevs, which corresponds to smaller physical Hubble volumes during the PT. In the right panel, we show the measures of the bubble radius at percolation, normalized to the Hubble length  (mean, approximated mean, and radius corresponding to the mode of the bubble volume distribution). As the inverse timescale of the transition does not need to vary much from $\beta_H \approx 8$ in order to give $f_{\PBH}=1$ over the entire allowed window, the typical bubble sizes when normalized to Hubble are also always similar.

\section{Gravitational waves}

\subsection{The spectra}

We next turn to the spectra we employ for the gravitational wave signal defined as
	\begin{equation}
	h^{2}\Omega_{\GW}(f) \equiv \frac{h^{2}}{ \rho_{c}} \frac{ d \rho_{\GW}}{ d \log f },
	\end{equation}
where $\rho_{\GW}$ is the energy density in GWs and $\rho_{c}$ is the critical density.

As spectra from dedicated studies are often given at production rather than today, we briefly review how to take into account the redshifting. We note that GWs redshift like radiation but are not reheated, and $\Omega_{\rm rad}(T_{\RH})  \equiv \rho_{\rm rad}/\rho_{c}  \simeq 1$ just after the PT. The above implies an amplitude today~\cite{Kosowsky:1992rz,Konstandin:2017sat} 
	\begin{subequations}
	\begin{align}
	h^{2}\Omega_{\GW} & = h^{2}\Omega_{\rm rad}(T_0) \left( \frac{ g_{\ast s}(T_0) }{ g_{\ast s}(T_{\RH}) } \right)^{4/3} \frac{ g_{\ast }(T_{\RH}) }{ g_{\ast}(T_0) } \Omega_{\mathrm{GW} \ast}  \\
	                  & = \frac{ g_{\ast s}(T_0)^{4/3} \pi^{2} T_0^{4}}{30  } \frac{8 \pi}{ 3 H_{100}^{2} M_{\Pl}^2}  \frac{   \Omega_{\mathrm{GW} \ast} }{ g_{\ast s}(T_{\RH})^{1/3} }  \\
		          & =  7.64 \times 10^{-5} \,  g_{\ast s}(T_{\RH})^{-1/3} \,  \Omega_{\mathrm{GW} \ast},
	\end{align}
	\end{subequations}
where $\Omega_{\mathrm{GW} \ast}$ is the GW signal just after the PT, $T_{0} \simeq 0.23$ meV is the CMB temperature today, and $H_{100} \equiv 100$~km/s/Mpc. The frequency of the signal is redshifted as
	\begin{equation}
	f_{0} = \left( \frac{ g_{\ast s}(T_0) }{ g_{\ast s}(T_{\RH}) } \right)^{1/3} \frac{ T_{0} }{ T_{\RH} } \, f_{\ast}.
	\end{equation}
where $f_{\ast}$  is the frequency just after the PT. Concretely, for the reheating temperature, we take 
	\begin{equation}
	T_{\RH} = \left( \frac{ 90M_{\Pl}^{2}H_{\rm bkg}(t_{p})^2 }{ g_{\ast}(T_{\RH})8\pi^3 } \right)^{1/4},
	\end{equation}
which follows directly from the first Friedmann equation.  We are interested in supercooled PTs with highly relativistic bubble walls. We will use three determinations of the spectrum, showing our conclusions are hardly affected by the precise choice.

\begin{figure}[t]
\begin{center}
\includegraphics[width=210pt]{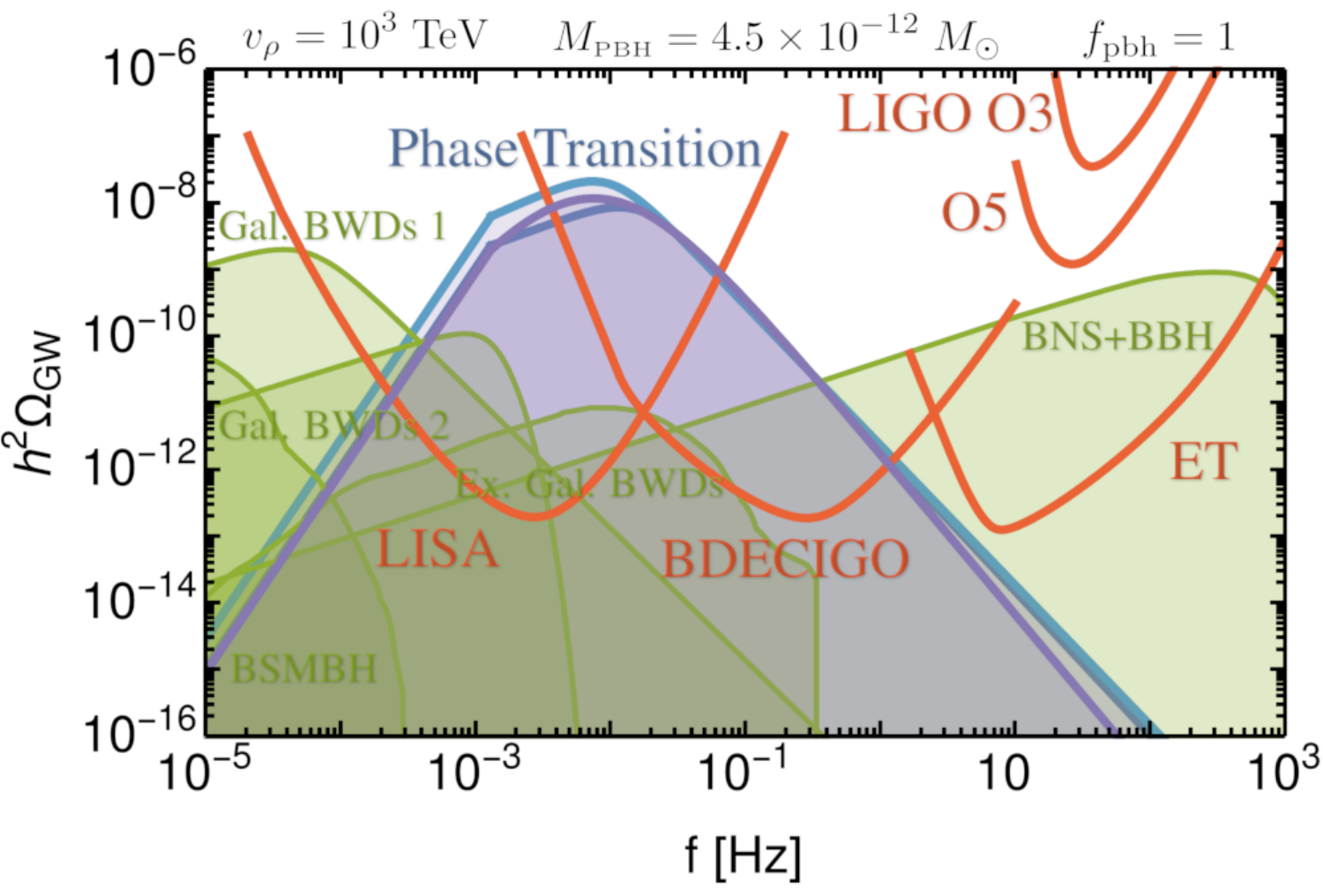} $\quad$
\includegraphics[width=210pt]{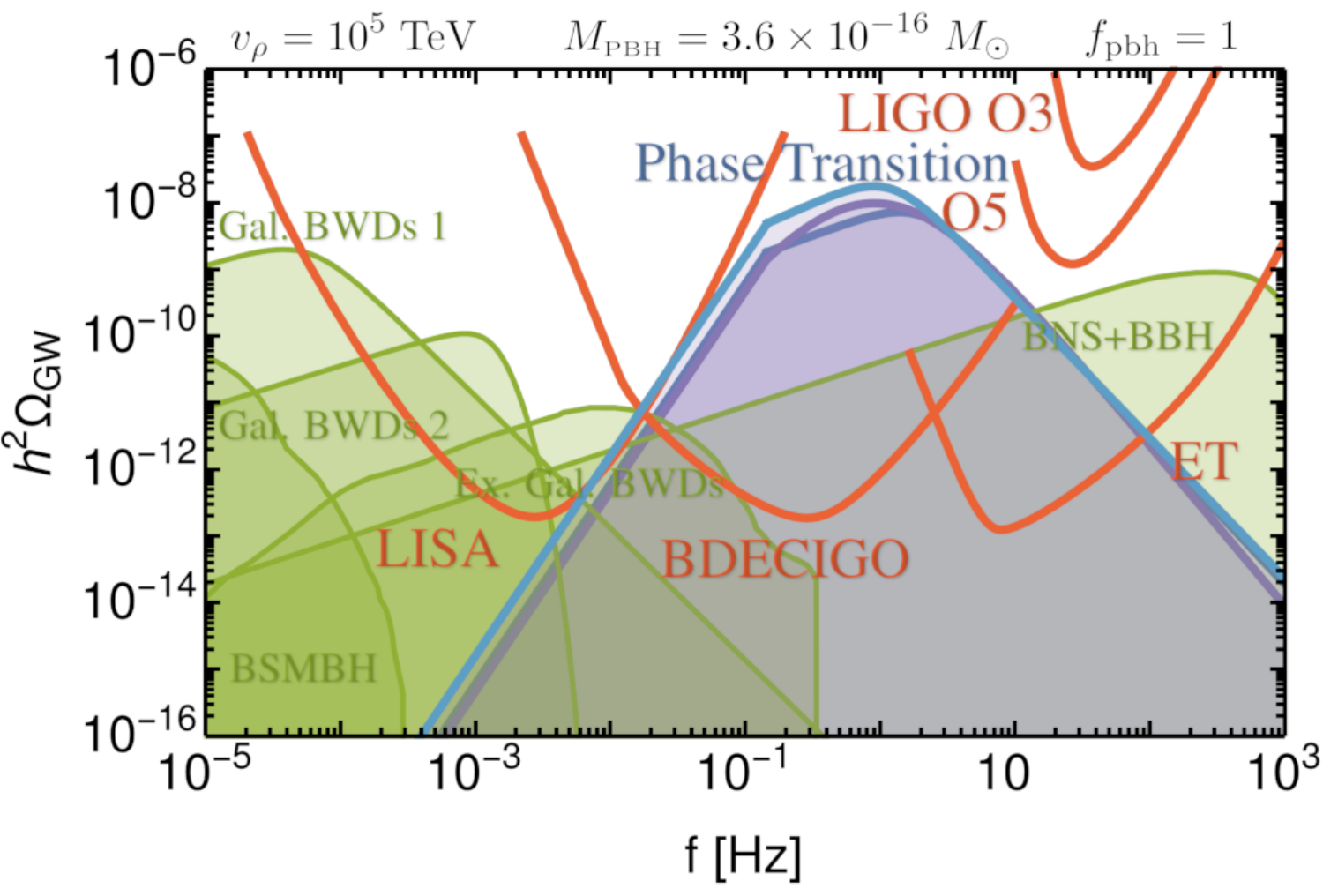}
\end{center}
\caption{\small GW spectra for two PTs returning $f_{\PBH}=1$. For each of the PTs, we show three estimates of the spectra --- given the macroscopic PT parameters --- available in the literature. Also shown are power-law-integrated sensitivity curves ($\mathrm{SNR}=10$) for current and future interferometers together with astrophysical foregrounds. Other key parameters associated with these example PTs can be read off Fig.~\ref{fig:gblreq}.}
\label{fig:GWSPECexample}
\end{figure}

\begin{itemize}
\item The first estimate comes from $(3+1)$ dimensional lattice simulations of thick wall bubble collisions by Cutting, Escartin, Hindmarsh, and Weir~\cite{Cutting:2020nla}. The spectrum is found to be
	\begin{equation}
	h^{2}\Omega_{\rm thw}(f) = 4.38 \times 10^{-9} \left( \frac{ 100 }{  g_{\ast }(T_{\RH}) } \right)^{1/3} \left( \frac{ 10 }{ \beta_{H} } \right)^{2}  S_{\rm thw}(f), 
	\end{equation} 
where the spectral shape is given by
	\begin{equation}
	S_{\rm thw}(f) =  \frac{2.902 \, \tilde{f}_{\rm thw}^{2.16} \, f^{0.742}}{2.16\tilde{f}_{\rm thw}^{2.902}+0.742f^{2.902}},
	\end{equation}
and the peak frequency is
	\begin{equation}
	\tilde{f}_{\rm thw} =  32 \; \mathrm{mHz} \, \left( \frac{ g_{\ast }(T_{\RH}) }{ 100 } \right)^{1/6} \left(  \frac{ \beta_H }{ 10 } \right)   \, \left( \frac{ T_{\RH} }{ 10^{2} \; \mathrm{TeV} } \right).
	\end{equation}
In the above, we have used the central values of the fit provided in~\cite{Cutting:2020nla} for the thickest walls of their simulations, and taken the bubble diameter to be $D_{\rm bub}H \simeq (8 \pi)^{1/3}\beta_{H}^{-1}$~\cite{Enqvist:1991xw} in converting the spectrum in \cite{Cutting:2020nla} to be in terms of $\beta_{H}$. From Fig.~\ref{fig:gblreq} we see this diameter is in good agreement with the mean diameter extracted from the bubble distribution when using $\beta_{H}(T_n)$, which we subsequently use in calculating the GW spectrum. Note if we were to instead use $2R_{\rm max}$, the diameter at which the bubble energy distribution is maximized, as advocated in~\cite{Ellis:2018mja}, our GW signals would be stronger.
\item The second estimate comes from the hybrid simulations by Vaskonen and Lewicki~\cite{Lewicki:2020azd}. Therein, the anisotropic stress induced in a bubble collision is first determined in a (1+1) dimensional simulation. This is then used as a source at points at which walls collide in a  $(3+1)$ dimensional lattice simulation in the thin walled limit. The advantage here is that the lower dimensional simulation allows one to study the effect of non-trivial scalar gradients and associated gauge field production during the bubble collision. The simulations find differences in the spectra between the non-gauged~\cite{Lewicki:2020jiv} and gauged cases~\cite{Lewicki:2020azd}, of U(1) symmetry breaking. Of course we use the gauged example here, where the spectrum is 
	\begin{equation}
	h^{2}\Omega_{\rm hyb}(f) = 5.93 \times 10^{-9} \left( \frac{ 100 }{  g_{\ast }(T_{\RH})} \right)^{1/3} \left( \frac{ 10 }{ \beta_{H} } \right)^{2}  S_{\rm hyb}(f),
	\end{equation} 
with the shape,
	\begin{equation}
	S_{\rm hyb}(f) = \frac{ 695 }{ \left[ 2.41 \left( \frac{f}{\tilde{f}_{\rm hyb}} \right)^{-0.557} + \left(  \frac{f}{\tilde{f}_{\rm hyb}} \right)^{0.574} \right]^{4.20} }
	\end{equation}
and peak frequency
	\begin{equation}
	\tilde{f}_{\rm hyb} =  22 \; \mathrm{mHz} \, \left( \frac{ g_{\ast} }{ 100 } \right)^{1/6} \left(  \frac{ \beta_H }{ 10 } \right)   \, \left( \frac{ T_{\RH} }{ 10^{2} \; \mathrm{TeV} } \right).
	\end{equation}
\item Reassuringly also semi-analytic methods which avoid the need for running lattice simulations have been developed~\cite{Jinno:2017fby,Konstandin:2017sat}. We use the results of the bulk flow from Konstandin~\cite{Konstandin:2017sat} with amplitude
	\begin{equation}
	h^{2}\Omega_{\rm bulk}(f) =  1.06 \times 10^{-8}  \left( \frac{ 100 }{ g_{\ast }(T_{\RH}) } \right)^{1/3} \left( \frac{ 10 }{ \beta_{H} } \right)^{2} S_{\rm bulk}(f),
	\end{equation}
and spectral shape
	\begin{equation}
	S_{\rm bulk}(f) = \frac{ 3\tilde{f}_{\rm bulk}^{2.1}f^{0.9} }{2.1\tilde{f}_{\rm bulk}^3+0.9f^{3}},
	\end{equation}
with peak frequency at
	\begin{equation}
	\tilde{f}_{\rm bulk} =  21 \; \mathrm{mHz} \, \left( \frac{  g_{\ast }(T_{\RH})}{ 100 } \right)^{1/6} \left(  \frac{ \beta_H }{ 10 } \right)   \, \left( \frac{ T_{\RH} }{ 10^{2} \; \mathrm{TeV} } \right).
	\end{equation}
\end{itemize}	
Note all the above results are given in the limit $\alpha \gg 1$, $v_{w} \simeq 1$, and the spectral shape functions have been normalized to return unity at their respective peak frequencies. As the finite cosmological horizon is not taken into account in the above simulations, we also impose the correct scaling $\Omega_{\GW} \propto f^{3}$, for super-horizon modes at the time of the PT~\cite{Durrer:2003ja,Caprini:2009fx,Barenboim:2016mjm,Cai:2019cdl,Hook:2020phx}. These correspond to the frequenices at IR tail of the spectrum, below
	\begin{equation}
	f^{\rm PT}_{\rm horizon} =  2.6 \; \mathrm{mHz} \, \left( \frac{  g_{\ast }(T_{\RH}) }{ 100 } \right)^{1/6} \, \left( \frac{ T_{\RH} }{ 10^{2} \; \mathrm{TeV} } \right), 
	\end{equation}
as measured today. In Fig.~\ref{fig:GWSPECexample} we show the resulting GW spectra for two example parameter points with $f_{\PBH} = 1$. For practical purposes here, we see the GW spectra from the three estimates are all rather similar. They are all above foregrounds and detectable at some upcoming interferometer. 

One word of caution is in order. The above estimates do not take into account the Hubble expansion during the transition itself. The issue has been studied in Ref.~\cite{Zhong:2021hgo}, which extended the analysis to an expanding background, but only for PTs in a radiation dominated background and using the older envelope approximation. The results indicated a suppression of an order-of-magnitude in the signal for PTs with $\beta_{H} \approx 10$. Note even with such a suppression, our GW signals would still be easily detectable, although this effect adds considerable theory uncertainty. To better gain a handle on the GW signal, the bulk flow analysis or the simulations will also have to be modified to take into account expansion, in particular for a vaccum dominated background. Our analysis further motivates such efforts.

\begin{figure}[p]
\begin{center} 
\includegraphics[width=210pt]{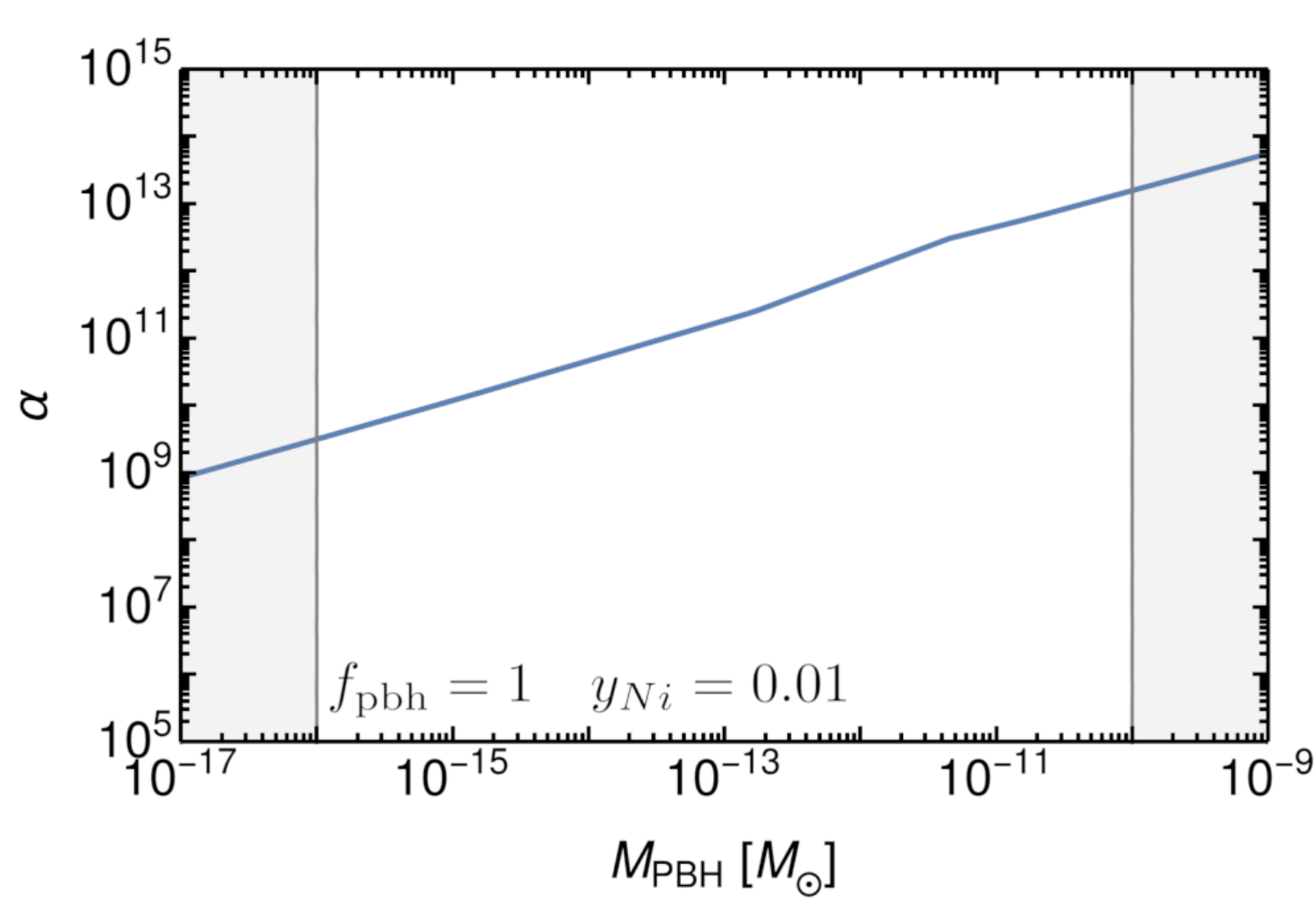} $\quad$
\includegraphics[width=210pt]{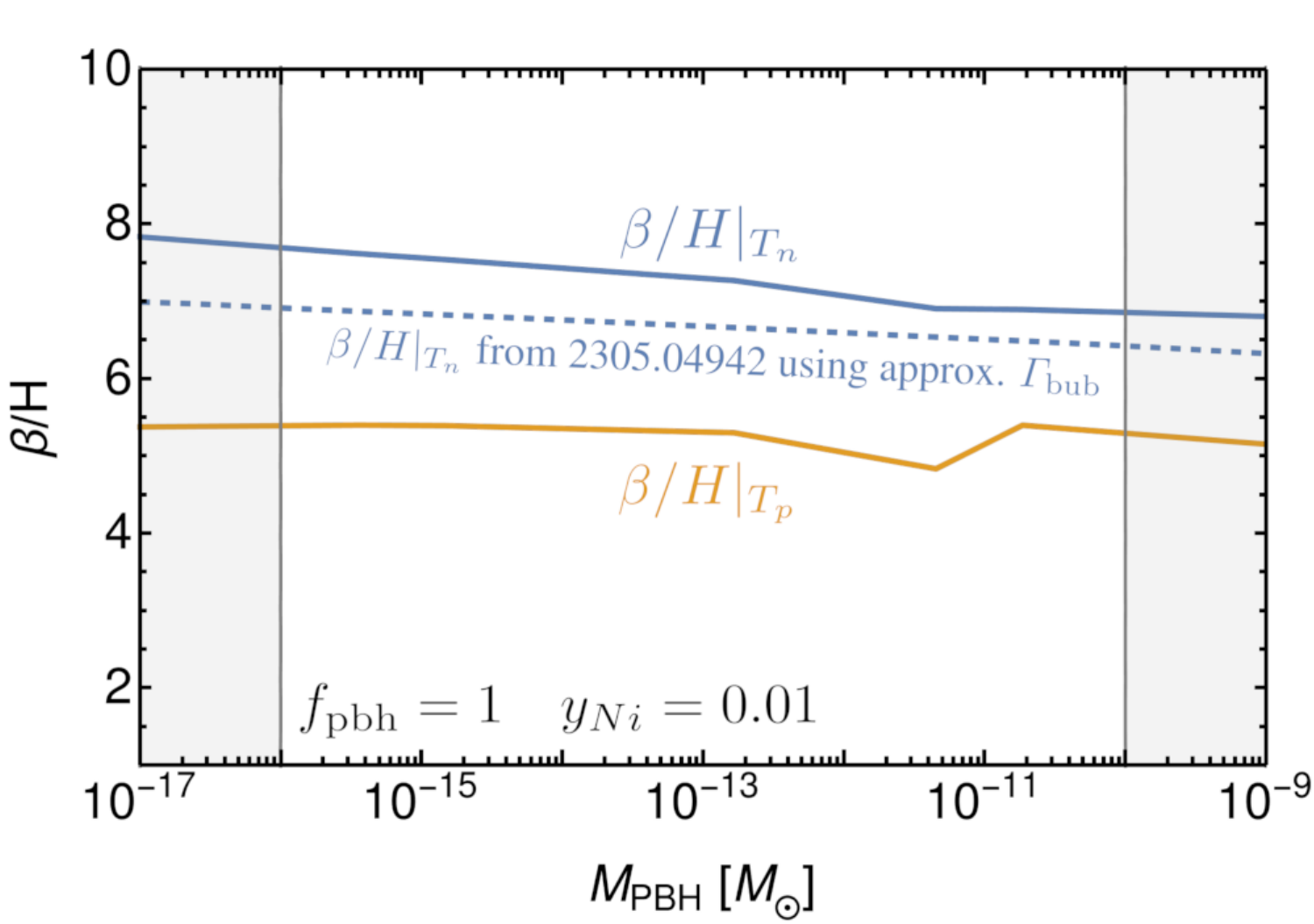}
\\
\includegraphics[width=210pt]{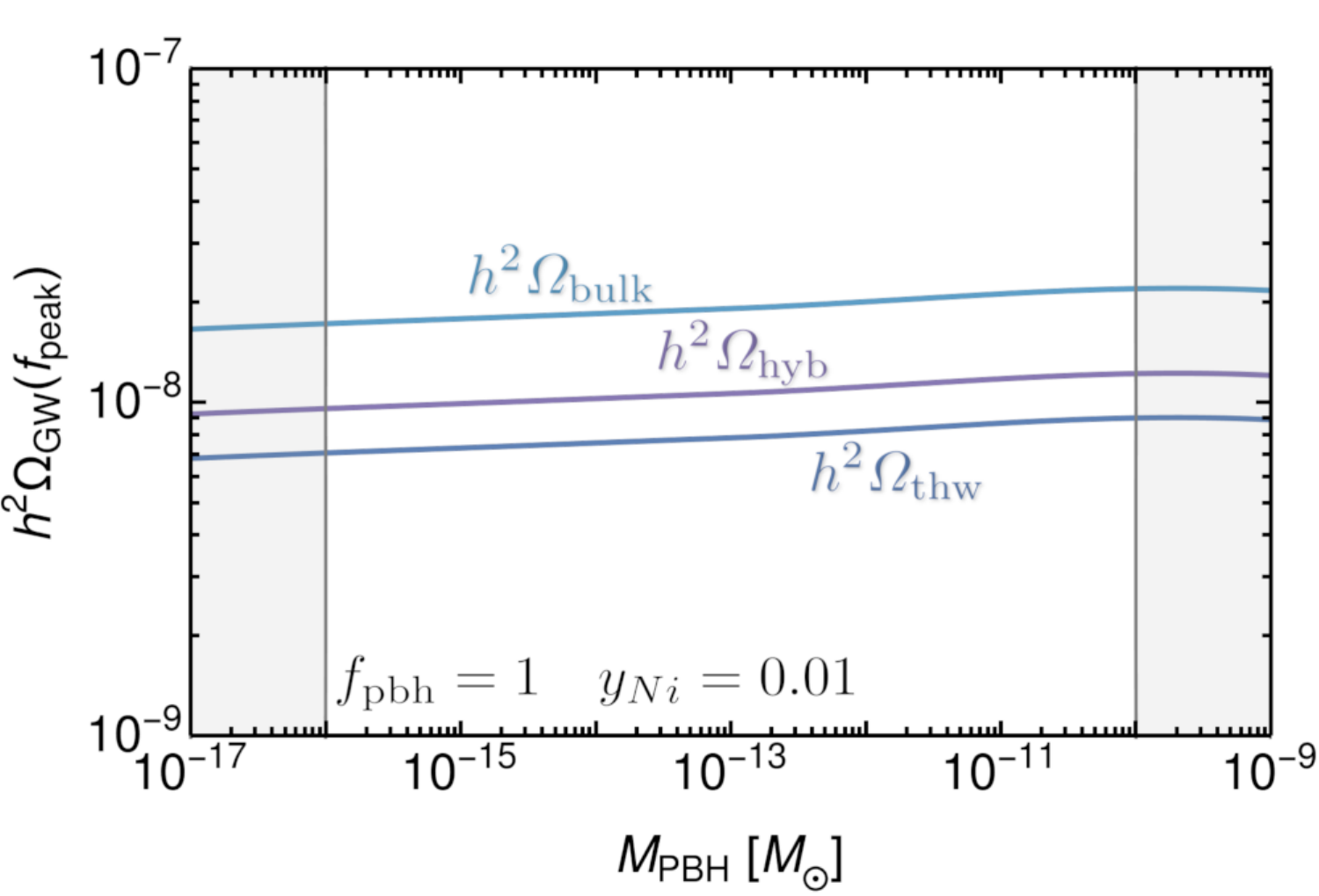} $\quad$
\includegraphics[width=210pt]{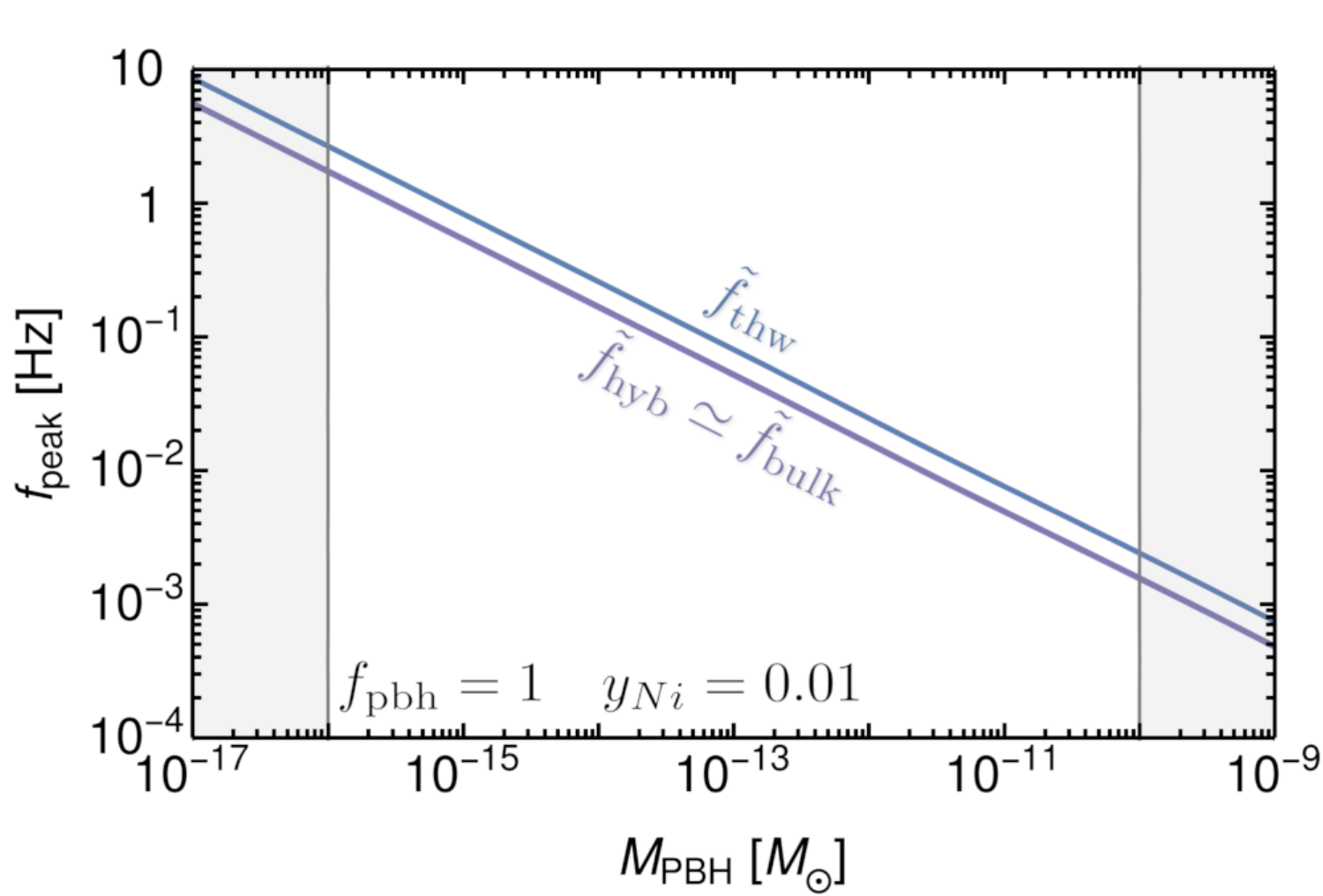}
\\
\includegraphics[width=210pt]{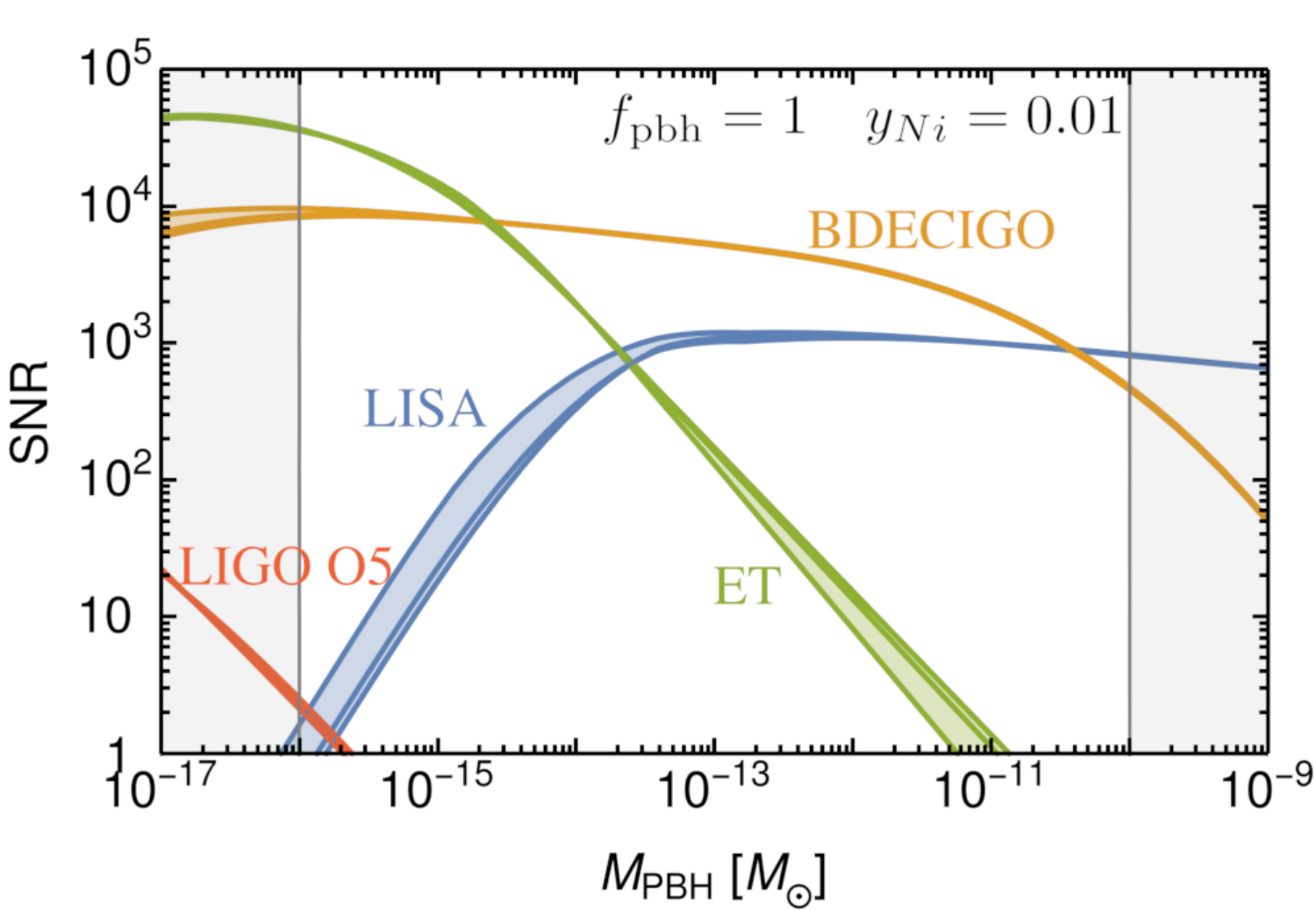} $\quad$
\includegraphics[width=210pt]{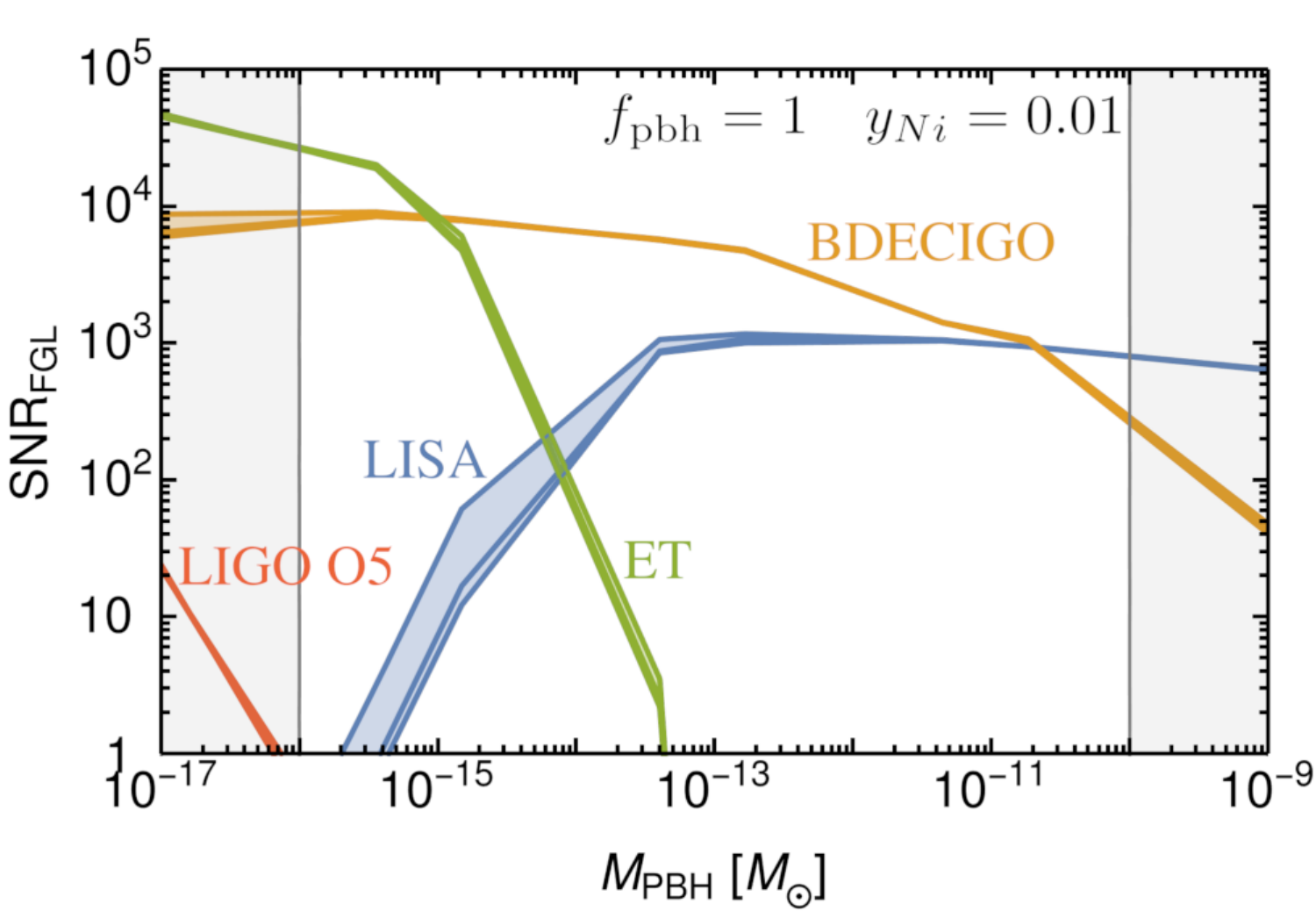}
\end{center}
\caption{\small Top: The latent heat $\alpha$ and inverse timescale of the transitions $\beta/H$ returning $f_{\PBH}=1$ as a function of $M_{\PBH}$. Also shown is a comparison to the results of~\cite{Gouttenoire:2023naa}, which uses the approximate $\Gamma_{\rm bub}$ ansatz, so some differences are expected. Middle: The peak amplitude and peak frequency of the three GW spectrum models as a function of $M_{\PBH}$. Bottom: the signal-to-noise ratio ($\mathrm{SNR}$), and signal-to-noise ratio above foregrounds ($\mathrm{SNR}_{\rm FGL}$), as a function of $M_{\PBH}$. The signal can be detected by BDECIGO or a combination of LISA and ET over the entire allowed range of $M_{\PBH}$ with $f_{\PBH}=1$. The shaded band encompasses the three different determinations of the GW signal. Additional theory and experimental forecasting uncertainty means the true error band is larger.}
\label{fig:SNR}
\end{figure}

\subsection{Signal-to-noise ratio}

We now wish to estimate the signal-to-noise ratio and also show foregrounds will not mask our signal. A similar method to what we describe in this section, up to some small updates, has also been used in~\cite{Baldes:2021aph}. The signal-to-noise ratio is given by~\cite{Allen:1997ad,Kudoh:2005as,Thrane:2013oya,Caprini:2019pxz,Brzeminski:2022haa}
	\begin{equation}
	\mathrm{SNR} = \sqrt{t_{\rm obs} \int \left(\frac{\Omega_{\GW}^2}{\Omega_{\rm sens}^2 + 2\Omega_{\GW}\Omega_{\rm sens}+	2\Omega_{\GW}^2}\right) df},
	\label{eq:snr}
	\end{equation}
where $t_{\rm obs}$ is the observation time and $\Omega_{\rm sens}$ the sensitivity of the interferometer. For these quantities we take without deviation the choices made in~\cite{Baldes:2021aph} and use this to also calculate the power-law-integrated sensitivity curves shown in Fig.~\ref{fig:GWSPECexample}.
Note the second and third terms in the denominator of the integral in Eq.~\eqref{eq:snr} means the SNR is saturated for large GW signals.
We do not want to count signals below astrophysical foregrounds as detectable, so as a simple and naive estimate, we define a foreground limited signal-to-noise ratio in which we only count GW signals above the astrophysical contamination. Namely, we compute,
	\begin{equation}
	\mathrm{SNR}_{\rm FGL} = \sqrt{t_{\rm obs} \int \left(\frac{\mathrm{Max} [0,\Omega_{\GW}(\nu)-\Omega_{\rm FG}(\nu)]^2}{\Omega_{\rm sens}^2 + 2\Omega_{\GW}\Omega_{\rm sens}+	2\Omega_{\GW}^2}\right) df}.
	\label{eq:snr}
	\end{equation}
As estimates of the foregrounds we take:
	\begin{itemize}
	\item  The component of the GW galactic white dwarf binaries which are not subtractable after four years of LISA observations, with approximate form given in~\cite{Cornish:2018dyw,Schmitz:2020rag}. The same quantity has also been estimated in~\cite{Lamberts:2019nyk}, with a higher peak value, but at lower frequencies. For this latter estimate we use the analytic fit from~\cite{Boileau:2022ter}. To be conservative we simply include both estimates.
	\item  The upper value of the extra-galactic white-dwarf binary estimate from~\cite{Farmer:2003pa}, a broadly similar estimate is derived in~\cite{Rosado:2011kv}. 
	\item  The central value binary neutron star and binary black hole signal inferred from the observations in~\cite{Abbott:2021xxi}, extrapolated to lower frequencies assuming the $f^{2/3}$ scaling, as is appropriate for an stationary ensemble of binaries with circular orbits losing energy solely through GW emission~\cite{Phinney:2001di}.
	\item The continuous super-massive binary black hole signal estimate of~\cite{Rosado:2011kv} will not mask our signal, but we display it in our plots.
	\end{itemize}
To summarize, in Fig.~\ref{fig:SNR} we show the latent heat $\alpha$, $\beta_{H}$, peak amplitude $ \Omega_{\GW}(f_{\rm peak})$, the peak frequency $f_{\rm peak}$, $\mathrm{SNR}$, and $\mathrm{SNR}_{\rm FGL}$ as a function of $M_{\PBH}$ (the vev $v_{\BL}$ and coupling are fixed by $M_{\PBH}$ and the requirement $f_{\PBH}=1$).

In the top row of Fig.~\ref{fig:SNR} we see, as expected, PBH formation occurs for $\alpha \gg 1$ and $\beta_{H}(T_{n}) \lesssim 8$, with slightly slower transitions required for more massive PBHs, in order to compensate the smaller enhancement in $\rho_{\PBH}/\rho_{\rm rad}$ between PBH formation time and matter-radiation equality. We also compare to the required $\beta_{H}$ found for $f_{\PBH}=1$ in Ref.~\cite{Gouttenoire:2023naa}, which used the approximate ansatz Eq.~\eqref{eq:gammaapprox} for the nucleation rate. We obtain the required abundance with a slightly higher $\beta_{H}$, because the bubble nucleation rate is somewhat suppressed in the full calculation compared to the approximate method. Note the collapse probability, $P_{\rm coll}$, is very sensitive to $\beta_{H}$, thus a small shift in the latter can compensate for changes in $\Gamma_{\rm bub}$ between the full and approximate expressions.

We thus confirm the approximate method is valid, at least in close-to-conformal potentials as studied here, which do not have a zero-temperature barrier. (If a zero temperature barrier exists, then there is the parameter dependent risk of having the field become permanently trapped in the false vacuum~\cite{Biekotter:2021ysx,Biekotter:2022kgf}. Then PBH formation would not occur simply because the background patches themeselves do not percolate. Such models may require a more careful examination. Nevertheless, bubble sizes and hence GW signals are still necessarily large in successully percolating models which produce PBHs and feature more dramatic decreases in $\Gamma_{\rm bub}$ below $T_{n}$, see~\cite{Lewicki:2023ioy} for work along these lines.)

In the middle row, we see the amplitude in $\Omega_{\rm GW}$ is large and almost constant over the range of $M_{\PBH}$, as $\beta_{H}$ is almost constant. The peak frequency covers three orders of magnitude, ranging over the sensitivies for LISA, BDECIGO, and ET, as anticipated through Eqs.~\eqref{eq:PBHmassapprox} and \eqref{eq:freqapprox}.

In the bottom row of Fig.~\ref{fig:SNR}, we observe that the typical supercooled phase transitions in the classically conformal $B-L$ model explaining all the dark matter in our universe in the form of PBHs will give rise to extremely strong GWs signals detectable by LISA, BDECIGO, and ET. These experiments will be able to probe the entire parameter space of the model producing $f_{\PBH} = 1$.  Current limits do not constrain this PBH formation mechanism in the given range of $M_{\PBH}$~\cite{KAGRA:2021kbb,Badger:2022nwo}.

\section{Validity of assumptions}
\subsection{Rapid decay of the condensate excitations}

\begin{figure}[t]
\begin{center} 
\includegraphics[width=200pt]{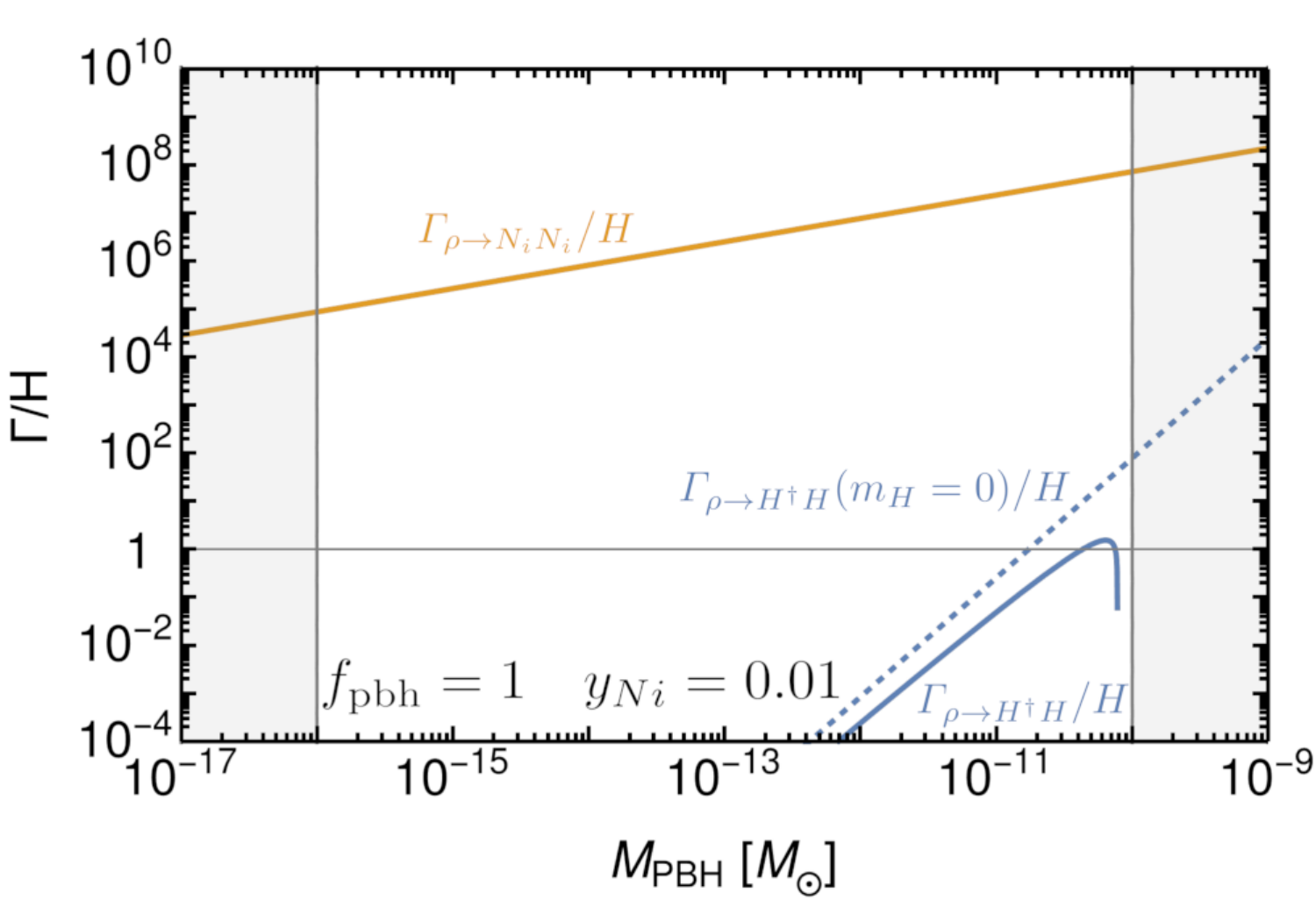} $\quad$ \includegraphics[width=200pt]{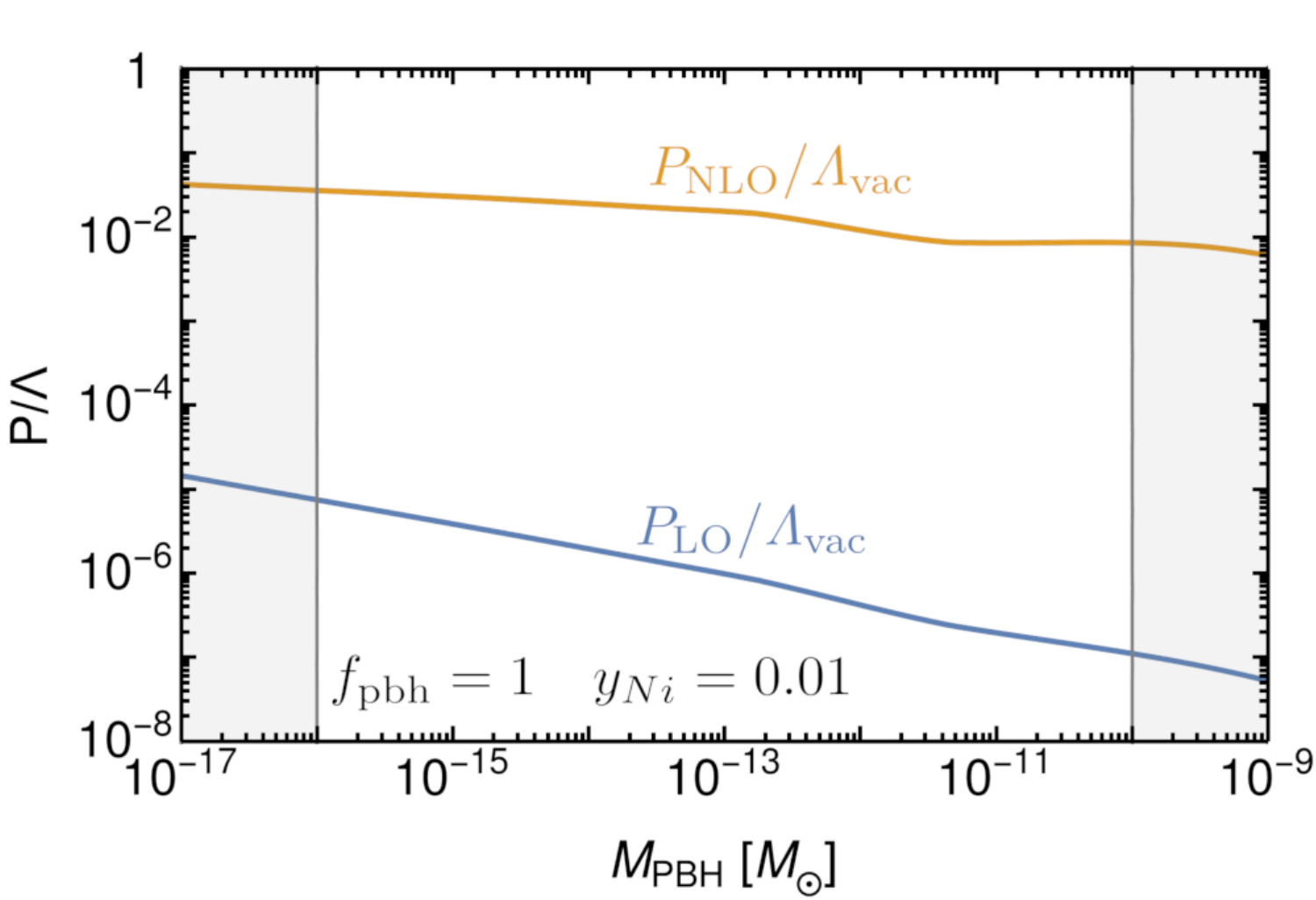} 
\end{center}
\caption{\small Left: The decay rates compared to the Hubble rate. Right: The LO and the $\gamma$ enhanced NLO retarding pressure on the wall compared to the driving pressure, i.e.~the vacuum energy difference.}
\label{fig:rates}
\end{figure}

In the above, we have assumed rapid reheating, so that $T_{\RH} \simeq T_{\rm infl}$. To check whether this is a valid assumption in our model, we can compare the decay rate of the scalar $\rho$ with the Hubble rate at the end of the PT. The first partial width we consider is of $\rho$ decaying to the EW Higgs doublet
	\begin{equation}
	\Gamma_{\rho \to H^{\dagger}H} = \frac{ \lambda_{\rho h}^{2} v_{\rho}^{2} }{ 8 \pi m_{\rho} } \, \mathrm{Re} \left[ \sqrt{ 1 - \frac{ 4m_{H}^2 }{ m_{\rho}^2 } } \, \right].
	\end{equation}	
Here we take into account the thermal mass of the Higgs, which begins to become present as the bath is reheated, 
	\begin{equation}
	m_{H}^{2} \approx \left( \frac{ \lambda_{h}^{2} }{ 2 } + 3 \frac{  g_{2}^{2} }{ 16 } +  \frac{ g_{Y}^{2} }{ 16 } + \frac{  \lambda_{h \rho} }{ 12 }  \right) T^{2} \approx 0.4 T^{2}.
	\end{equation}
The above width is tiny, due to the suppression of $\lambda_{\rho h}^{2}$, and possible kinematic factors as $m_{H}(T)$ becomes large compared to $m_{\rho}$.
But we can also consider the partial width to the $N_{i}$, 
	\begin{equation}
	\Gamma_{\rho \to N_iN_i} = \frac{ y_{Ni}^2 m_{\rho} }{ 32 \pi } \left( 1 - \frac{2 M_{Ni}^2 }{ m_{\rho}^2 } \right)\, \mathrm{Re} \left[ \sqrt{ 1 - \frac{ 4M_{Ni}^2 }{ m_{\rho}^2 } } \, \right].
	\label{eq:dectoN}
	\end{equation}
As $M_{Z'} \approx 10T_{\RH}$, the thermal masses of the $N_{i}$ are negligible following the PT. For standard assumptions regarding the typical Yukawa coupling size --- as inferred in from the type-I seesaw and the atmospheric mass squared difference of the active neutrinos --- we can assume the $N_{i}$ decay rapidly compared to the Hubble rate, i.e.~the strong washout regime~\cite{Buchmuller:2004nz}. We can thus simply compare Eq.~\eqref{eq:dectoN} to the Hubble rate to determine whether we can avoid a long lived $\rho$ or other matter domination following the PT. In Fig.~\ref{fig:rates} left we show both the partial widths compared to the Hubble rate for our parameter choice, showing our assumptions of rapid decay, giving $T_{\RH} \simeq T_{\rm infl}$ is justified. Note the $Z'$ will also rapidly decay to SM fermions.

\subsection{Pressure on the bubble wall}

For supercooled transitions, such as those studied here, leading order pressure on the wall as particles gain their mass~\cite{Bodeker:2009qy},
	\begin{equation}
	P_{\LO} \approx \frac{T_{p}^{2}}{24} \left( 3M_{Z'}^2 + m_{\rho}^{2} +  M_{Ni}^{2}  \right),
	\end{equation}
is generically insufficient to stop the wall from accelerating. The presence of gauge bosons which change their mass across the wall, however, leads to a NLO retarding pressure which increases linearly with  $\gamma$, the bubble wall Lorentz factor~\cite{Bodeker:2017cim}. We therefore check whether the vacuum energy density is converted into the wall energy, or whether sufficient soft-gauge bosons are produced at the wall, for the energy to instead be stored in the latter before wall collision. Such a situation has been modelled in~\cite{Jinno:2019jhi}. Whether this would affect the GW signal appreciably is a separate question, macroscopically both cases seem to be captured by the bulk flow model (e.g.~see~\cite{Jinno:2019jhi} and discussion in~\cite{Baldes:2023fsp}). Soft gauge boson production leads to a retarding pressure on the wall $P_{\NLO} \propto \gamma T_{p}^{3} M_{Z'}$~\cite{Bodeker:2017cim,Azatov:2020ufh}. Here we use the leading-log determination for this term given in~\cite{Gouttenoire:2021kjv},
	\begin{equation}
	P_{\NLO} \approx  \frac{ \kappa \zeta(3)  (Q_{\BL}^{\rm eff})^2 \alpha_{\BL} \gamma M_{Z'} T_{p}^{3} }{ \pi^{3} } \log\left( \frac{ v_{\rho} }{ T_p } \right),
	\end{equation}
where $\kappa \approx 4$, $(Q_{\BL}^{\rm eff})^2 =40$ is an effective charge factor summing over the bath degrees-of-freedom with 3/4 (1) weighting for fermions (scalars), $\alpha_{\BL} \equiv g_{\BL}^2/4\pi$, and we have used the thermal mass of the $Z'$ as an IR cutoff in the logarithm.
The maximum Lorentz factor is obtained at collision for an effectively run-away wall, i.e.~approximately zero retarding pressure, and is given by~\cite{Gouttenoire:2021kjv}
	\begin{equation}
	\gamma \approx \frac{1}{3} \frac{ R_{\rm coll} }{ R_{\rm nuc} } \approx \frac{1}{3} \frac{\pi^{1/3}}{\beta_{H}H} \left( \frac{ T_{n} }{10} \right),
	\end{equation}
where $R_{\rm nuc}\approx 10/T_{n}$ is the bubble radius at nucleation and $R_{\rm coll} \approx \pi^{1/3}/\beta$ is the bubble radius at collision (we have checked both analytic estimates through numerical determination). For consistency of the above $\gamma$ estimate we require $P_{\rm NLO} < \Lambda_{\rm vac}$. We calculate these quantities and display the results in Fig.~\ref{fig:rates} right, showing the wall is in the effective run-away regime at collision.

\subsection{Reliability of the effective potential}

We now briefly discuss the reliability of the approximations we have made in deriving $V_{\rm eff}$. To answer this question requires the application of more refined techniques such as the renormalization group improved effective potential~\cite{Prokopec:2018tnq,Croon:2020cgk}. Indeed, in the context of the particle-DM dilution mechanisms, there are known examples in which going to the RGE improved $V_{\rm eff}$ changes the qualitative outcome of the mechanism itself~\cite{Kierkla:2022odc}. This is due to changes to the details of the nucleation, percolation, and reheating temperatures, although the PTs remain very strong. Such an investigation is left for future work.

\section{Comments on leptogenesis}

\begin{figure}[t]
\begin{center} 
 \includegraphics[width=240pt]{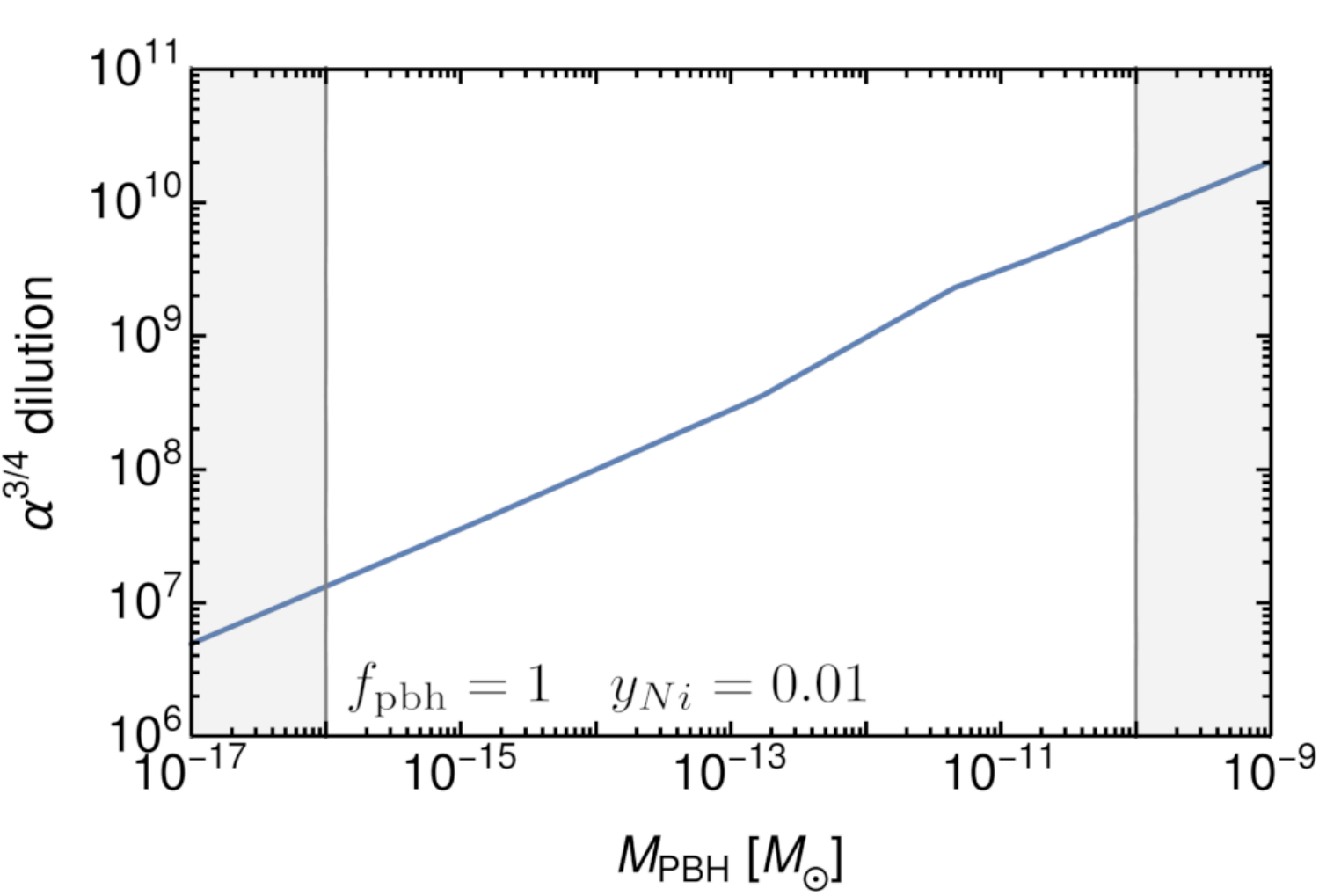} 
\end{center}
\caption{\small The entropy production (dilution) factor following the PT.}
\label{fig:dilution}
\end{figure}

In Fig.~\ref{fig:dilution} we show the dilution factor due to the entropy production following the PT. As the dilution factor is quite large, $\sim 10^{7}-10^{9}$, it seems unlikely that we could use a mechanism which relies on the dilute false vacuum plasma as a source of the baryon asymmetry.  For example, the mass gain mechanism when the $N_{i}$ suddenly gain their mass during the PT and then decay to generate the BAU~\cite{Baldes:2021vyz,Huang:2022vkf,Dasgupta:2022isg,Borah:2022cdx,Chun:2023ezg}. Similar conclusions most likely also hold for the related mechanism in~\cite{Azatov:2021irb}. Electroweak style baryogenesis mechanisms suffer from the same problem in the current context, together with a completely negligible yield for $v_{w} \simeq 1$, due to lack of diffusion back into the symmetric phase~\cite{Cline:2020jre}.

If instead, heavy particles are produced in the reheating process itself, following along the lines of~\cite{Masiero:1992bv,Katz:2016adq}, then these may have a larger abundance compared to the entropy density, and act as a source of the asymmetry. But note our bubble collisions are expected to be inelastic, which suppresses heavy particle production~\cite{Falkowski:2012fb}.

Anyway, for our choice of example Yukawa coupling $y_{i} =0.01$, the asymmetry will be generated by the usual thermal leptogenesis around $T \approx M_{N}$ after the PT (for our parameter choices $T_{\RH} \simeq 10 M_{Ni}$). In either case, resonant leptogenesis is generically required~\cite{Flanz:1994yx,Flanz:1996fb,Pilaftsis:1997jf,Pilaftsis:2003gt}, as the heavy neutrinos are much below the Davidson-Ibarra bound~\cite{Davidson:2002qv} (also see~\cite{Hambye:2001eu}). Note the heavy neutrinos are efficiently produced in the early universe due to their Yukawa coupling to $\rho$ and their gauge interactions.

\section{Conclusion}

In this work we have studied the GW signal from PBH formation during a PT. We have assumed $f_{\PBH}=1$, which limits the range for the peak frequency of the GWs, and calculated the required underlying parameters of the model to obtain the PBH abundance. We used the properties of the PT to find the expected GW signal from bubble collisions over the allowed parameter space, showing it can be detected by BDECIGO or a combination of LISA and ET, independently of the precise $M_{\PBH}$, over the entire range allowed for $f_{\PBH}=1$. Substituting one of the experiments by a broadly equivalent one will, of course, not affect the conclusions. An important caveat is that the GW predictions for strong phase transitions may overestimate the signal because expansion during the transition is not taken into account. This motivates further refinement of such estimates, along the lines of~\cite{Zhong:2021hgo}, but extended to vacuum dominated transitions and beyond the envelope approximation. We note a suppression in $\Omega_{\rm GW}$ by an order of magnitude for $\beta_{H} \approx 10$, would still return SNRs of $10^{2}$ and above over the $f_{\PBH}=1$ parameter space, so prospects remain very promising.

The classically scale invariant $B-L$ model we considered does not contain an automatically stable DM candidate, although given a small enough Yukawa coupling, $N_{1}$ could play the role of the possibility of a $\sim$ keV scale sterile neutrino (provided an appropriate mechanism was also in play in order to set its relic abundance correctly, as it carries gauge charge). Such sterile neutrino scenarios are by now by now heavily constrained~\cite{Boyarsky:2018tvu}. Additional field content can of course be included to act as DM~\cite{Arcadi:2023lwc}. We have instead proposed using the PBH generation mechanism from the strong PT realised with the close-to-conformal potential to obtain the observed $\Omega_{\rm DM}$. Leptogenesis in this model should occur sometime after the PT, through resonant enhancement of the CP violation, as in the usual low-scale thermal leptogenesis~\cite{Flanz:1994yx,Flanz:1996fb,Pilaftsis:1997jf,Pilaftsis:2003gt}. Furthermore, it would also be of interest to re-examine models of particle DM, in which the relic abundance relies on supercooled PTs~\cite{Hambye:2018qjv,Baldes:2018emh,Baldes:2020kam,Baldes:2021aph,Kierkla:2022odc,Wong:2023qon}, to check compatibility with PBH production. 

Our calculations in a specific model, taking into account the full temperature dependence of the bubble nucleation rate, have shown only modestly small changes to the inverse timescale of the transtion $\beta_{H}$ required to achieve $f_{\PBH}=1$, compared to model independent calculation which used an approximate ansatz for the nucleation rate. As $\beta_{H}$ sets the typical bubble size at collision and therefore also controls the properties of the GW signal, our results regarding the sensitivities of future detectors to such a late patch mechanism are independent of the underlying particle physics model.

Models of slow roll inflation producing PBHs also result in detectable GWs, in the same frequency range as the PT, due to the enhanced power spectrum on small scales leading to significant anisotropic stress~\cite{Nakama:2016gzw,Garcia-Bellido:2017aan,Cai:2018dig,Bartolo:2018evs,Bartolo:2018rku,Qin:2023lgo}. The additional power also present on small scales from the PT, would also induce anisotropic stress and a further GW source. Accordingly, this paper justifies the further development of computational techniques in order to pin down the expected PBH spectrum in the late patch mechanism (beyond the monochromatic approximation primarily used here), the bubble collision signal, together with any additional GWs at second order in perturbation theory, in light of testing $f_{\PBH}=1$ in upcoming interferometers. Along with the GW signal, the PBHs can be searched for using improved lensing studies, through future MeV telescopes~\cite{Ray:2021mxu,Agashe:2022jgk}, and the 21-cm absorption signal~\cite{Clark:2018ghm,Saha:2021pqf}.

%%%%%%%%%%%%%%%%%%%%%%%%%%%%%%%%%

\medskip

\section*{Acknowledgements}
We thank Yann Gouttenoire, Thomas Konstandin, Florian K\"uhnel, and Filippo Sala for useful discussions.

\paragraph*{Funding information}
This work was supported by the European Union’s Horizon 2020 research and innovation programme under grant agreement No 101002846, ERC CoG ``CosmoChart."

\medskip

\appendix

\section{Towards a non-monochromatic mass distribution}
\label{sec:nonmono}

In the main text, we used a monochromatic approximation for the PBH mass,
	\begin{equation}
	M_{\PBH} = \frac{4 \pi (\rho_{\rm rad late}+\rho_{\rm vac late}) c_{s}^{3}}{3 H_{\rm late}^{3}} = \frac{c_{s}^{3}}{2} \frac{ M_{\rm Pl}^2}{ H_{\rm late} }. \label{eq:appmassmonochromatic}
	\end{equation}
We also found 
	\begin{equation}
	f_{\PBH } = \frac{ \Omega_{\DM} + \Omega_{\rm B} }{ \Omega_{\DM} } \frac{ g_{\ast s}(T_{\rm eq}) }{ g_{\ast }(T_{\rm eq}) } \frac{ T_{\rm bkg \; form} }{ T_{\rm eq} } \frac{ \rho_{\PBH }(T_{\rm bkg \; form}) }{  \rho_{\rm rad }(T_{\rm bkg \; form}) },
	\end{equation}
where $(\Omega_{\DM} + \Omega_{\rm B})/\Omega_{\DM} \simeq 1.2$. We then used 
	\begin{equation}
	\frac{ \rho_{\PBH }(T_{\rm bkg \; form}) }{ \rho_{\rm rad }(T_{\rm bkg \; form}) }  \simeq  P_{\rm coll} \frac{ M_{\PBH} }{ M_{\rm Hor} }  \simeq   c_{s}^{3} P_{\rm coll}  \frac{ H_{\rm bkg}(t_{\delta \rm max}) }{ H_{\rm late}(t_{\delta \rm max}) }  , \label{eq:appmonochromatic}
	\end{equation}
 where $M_{\rm Hor} = M_{\Pl}^2/(2 H_{\rm bkg})$ is the energy inside the Hubble horizon of a background patch. In the monochromatic approximation $P_{\rm coll}(T_{i})$, given in Eq.~\eqref{eq:Pcoll}, is evaluated at $T_{i}$ chosen so that $\delta_{\rm max} = \delta_{c}$.
 
 We now assume instead critical collapse, as is known from numerical codes relevant for the inflation scenario, in which for $\delta_{m} > \delta_{c}$ the PBH mass is found to be~\cite{Choptuik:1992jv,Franciolini:2021nvv}
	 	\begin{equation}
	 	 M_{\PBH} = \mathcal{K} M_{\rm Hor} (\delta_{m} - \delta_{c})^{\gamma_c}, \label{eq:critcollapse}
	 	\end{equation}
where  $1 \lesssim \mathcal{K} \lesssim 10$ is an efficiency factor, and $\gamma_c \simeq 0.36$ is the critical exponent assuming radiation domination (both found from simulations). In the inflationary scenario, $\delta_{m}$ is the overdensity as measured at the maximum of the compactification function, at cosmological horizon crossing time, e.g.~see~\cite{Musco:2018rwt}. We now make assume something similar holds in the PT mechanism and replace $\delta_{m} \to \delta_{\rm max}$ in the above equation.  We warn we are making a big leap making such an assumption for critical collapse in the PT mechanism, and do so simply out of curiousity regarding the resulting PBH spectrum, mainly as an academic exercise. Note in using Eq.~\eqref{eq:critcollapse} we also do not include the small correction taking into account the different Horizon masses of the collapsing and background patches as we have done for our monochromatic approximation in Eq.~\eqref{eq:appmassmonochromatic}.

\begin{figure}[t]
\begin{center} 
 \includegraphics[width=200pt]{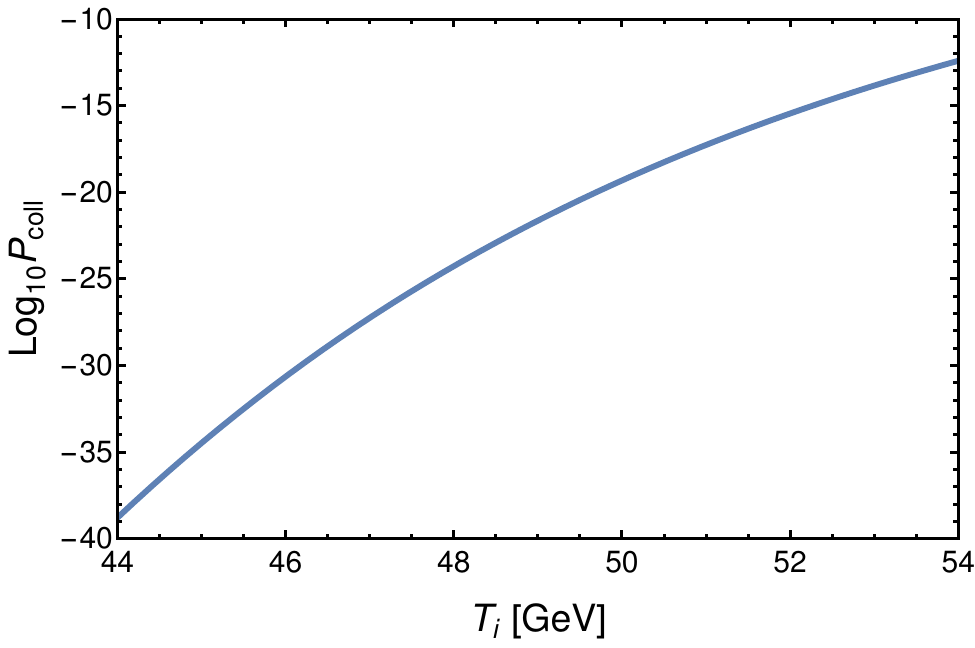} \quad  \includegraphics[width=200pt]{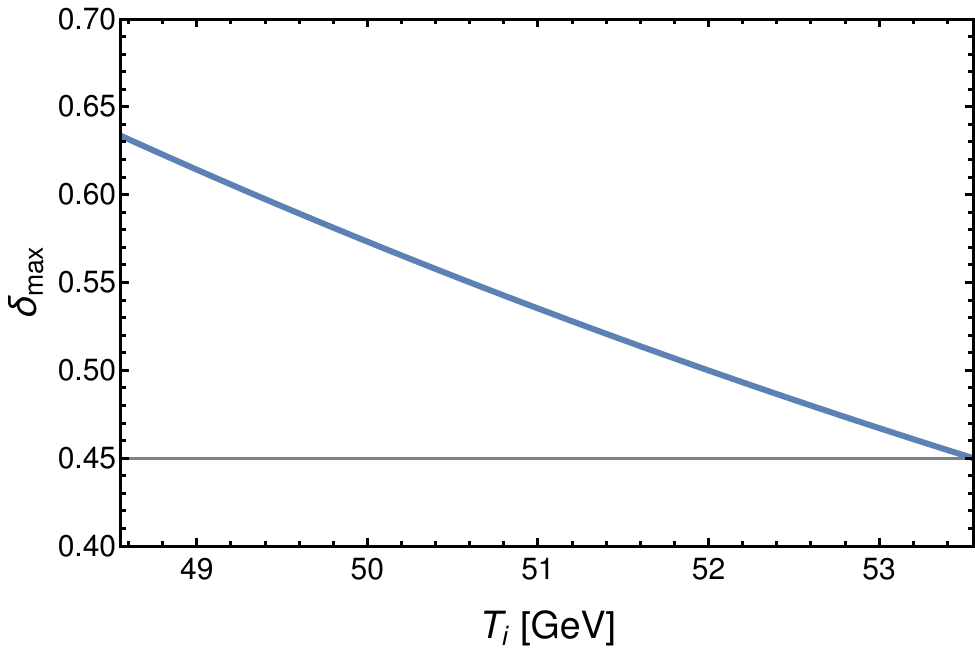}
\end{center}
\caption{\small Left: The cumulative probability $P(T_{i})$ for the example parameter point, $v_{\rho} = 10^{3}$ TeV, $y_{Ni} = 0.01$, and $g_{\BL} = 0.2958$. Right: For convenience we again show $\delta_{\rm max}(T_i)$ for the same parameter point, where $\delta_{\rm max}=0.45$ for $T_{i}=53.6$ GeV (for a larger range of $T_i$ see Fig.~\ref{fig:densities}, middle row, right).}
\label{fig:Pcoll}
\end{figure}

From Eq.~\eqref{eq:critcollapse} we see that patches with $\delta_{\rm max} = \delta_{c}$ fail to form a black hole, but instead a whole spectrum of PBH masses is generated from patches with larger $\delta_{\rm max}$, coming from progressively smaller $T_{i}$. The probability of keeping large regions bubble free also falls as $T_{i}$ is reduced. (Roughly speaking, $P_{\rm coll}$ is the probability of keeping a sufficiently large region bubble free sufficiently long, it therefore includes the probability of staying bubble free even longer. In other words it acts as a cumulative probability function.) A plot of $P_{\rm coll}(T_i)$ is shown in Fig.~\ref{fig:Pcoll}.

We now wish to calculate the resulting PBH mass spectra assuming Eq.~\eqref{eq:critcollapse} holds. The first spectrum typically considered in the literature is normalized to the PBH density and is conventionally defined as~\cite{Franciolini:2021nvv}
 	\begin{equation}
 	\Psi(M_{\PBH}) \equiv \frac{1}{\rho_{\PBH}} \frac{ d\rho_{\PBH} }{ dM_{\PBH} }.
 	\end{equation}
 In our scenario we have
	\begin{equation}
	\Psi(M_{\PBH})   = \frac{M_{\PBH}}{\mathcal{N}} \frac{ d P_{\rm coll} }{d M_{\PBH} } = \frac{M_{\PBH}}{\mathcal{N}} \frac{ d P_{\rm coll} }{ dT_i } \frac{ dT_i }{ d M_{\PBH}}  
		   = \frac{1}{ \mathcal{N} \gamma_c } \left( \frac{ M_{\PBH} }{ \mathcal{K} M_{\rm Hor } } \right)^{1/\gamma_c} \frac{ dP_{\rm coll} }{ dT_i } \frac{ dT_i }{ d\delta_{\rm max} }, \label{eq:nonmonochrommspectrum1}  
	\end{equation}
where the normalization factor is given by
	\begin{equation}
	\mathcal{N} = \int_{0}^{\infty} dM_{\PBH}' M_{\PBH}' \frac{ dP_{\rm coll} }{ dM_{\PBH}'}.
	\end{equation}
Note $dP_{\rm coll}/dM_{\PBH} < 0$ given our definition of $P_{\rm coll}$ in Eq.~\eqref{eq:Pcoll}, but the factor appears both in the numerator and the denominator, $\mathcal{N}$, so $\Psi$ is positive.
Integrating $\Psi(M_{\PBH})$ over the entire PBH mass range yields unity. In the above, we have related the spectrum to the collapse probability calculated in Eq.~\eqref{eq:Pcoll}, and we have used that $M_{\rm Hor}$ and $T_{\rm bkg \; form}$ are approximately independent of $M_{\PBH}$ for PBHs originating from overdensities with a preferred length scale (in the current context of the late patch mechanism this is simply the Hubble length during the PT). This is further justified by the strongly falling $P_{\rm coll}$ with decreasing $T_{i}$, as shown in Fig.~\ref{fig:Pcoll}.

The second commonly used spectrum is instead normalized to the observed DM density and is defined as~\cite{Franciolini:2021nvv}
	\begin{equation}
	f(M_{\PBH})  \equiv \frac{M_{\PBH}}{\rho_{\DM}} \frac{ d \rho_{\PBH} }{ dM_{\PBH} },
	\end{equation}
so that
	\begin{equation}
	f_{\PBH} = \int_{0}^{\infty} dM_{\PBH} \frac{f(M_{\PBH})}{M_{\PBH}} \qquad \text{and} \qquad f(M_{\PBH}) = f_{\PBH} M_{\PBH}\Psi(M_{\PBH}). 
	\end{equation}
In our scenario we find
	\begin{subequations}
	\begin{align}
	f(M_{\PBH}) & =  -\frac{ \Omega_{\DM} + \Omega_{\rm B} }{ \Omega_{\DM} }\frac{ g_{\ast s}(T_{\rm eq}) }{ g_{\ast }(T_{\rm  eq}) } \frac{ T_{\rm bkg \; form} }{ T_{\rm eq} } \frac{ M_{\PBH}^2 }{ M_{\rm Hor } } \frac{ dP_{\rm coll} }{ dM_{\PBH} }  \\
							& =   -\frac{ \Omega_{\DM} + \Omega_{\rm B} }{ \Omega_{\DM} } \frac{ g_{\ast s}(T_{ \rm eq}) }{ g_{\ast }(T_{ \rm eq}) } \frac{ T_{\rm bkg \; form} }{ T_{\rm eq} } \frac{ M_{\PBH}^2 }{ M_{\rm Hor } } \frac{ dP_{\rm coll} }{ dT_i } \frac{ dT_i }{ dM_{\PBH} } \label{eq:nonmonochromfspectrum1} \\
							& =   -\frac{ \Omega_{\DM} + \Omega_{\rm B} }{ \Omega_{\DM} } \frac{ g_{\ast s}(T_{ \rm eq}) }{ g_{\ast }(T_{ \rm eq}) } \frac{ T_{\rm bkg \; form} }{ T_{\rm eq} } \frac{1}{\gamma_c \mathcal{K}^{1/\gamma_c} } \left( \frac{ M_{\PBH} }{ M_{\rm Hor } } \right)^{1+1/\gamma_c} \frac{ dP_{\rm coll} }{ dT_i } \frac{ dT_i }{ d\delta_{\rm max} }, \label{eq:nonmonochromfspectrum2}
	\end{align}
	\end{subequations}
where we have again used that  $M_{\rm Hor}$ and $T_{\rm bkg \; form}$ are approximately independent of $M_{\PBH}$ for PBHs originating from overdensities with a preferred length scale. We now include an overall negative sign because $dP_{\rm coll}/dM_{\PBH} < 0$. For either of the spectra of interest, we can evaluate the last two differential factors numerically as functions of $M_{\PBH}$, and thus find the desired quantity.

\begin{figure}[t]
\begin{center} 
 \includegraphics[width=200pt]{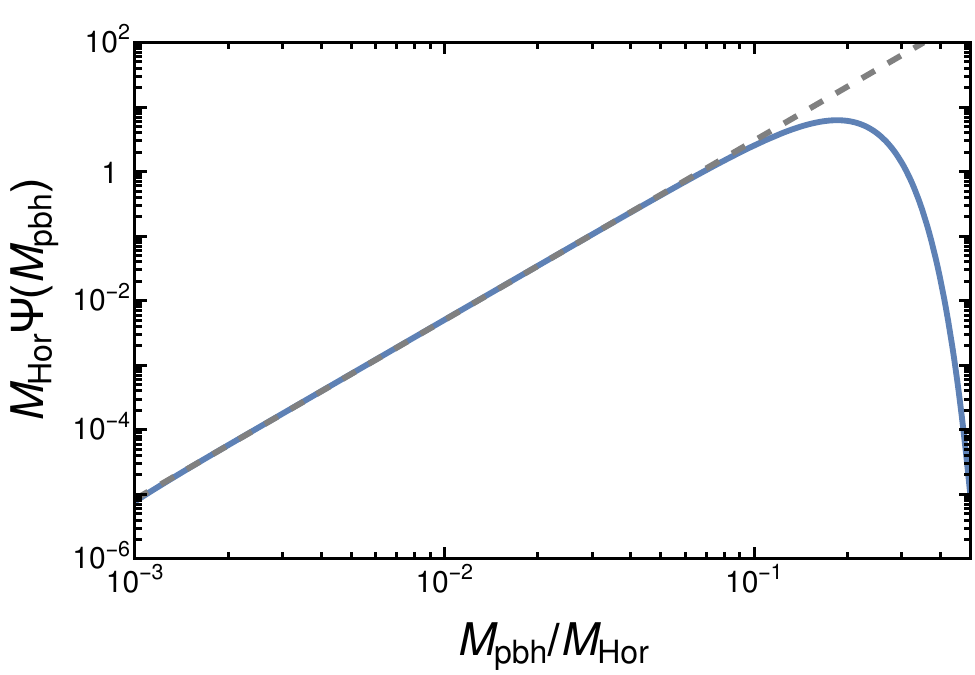} \quad \includegraphics[width=200pt]{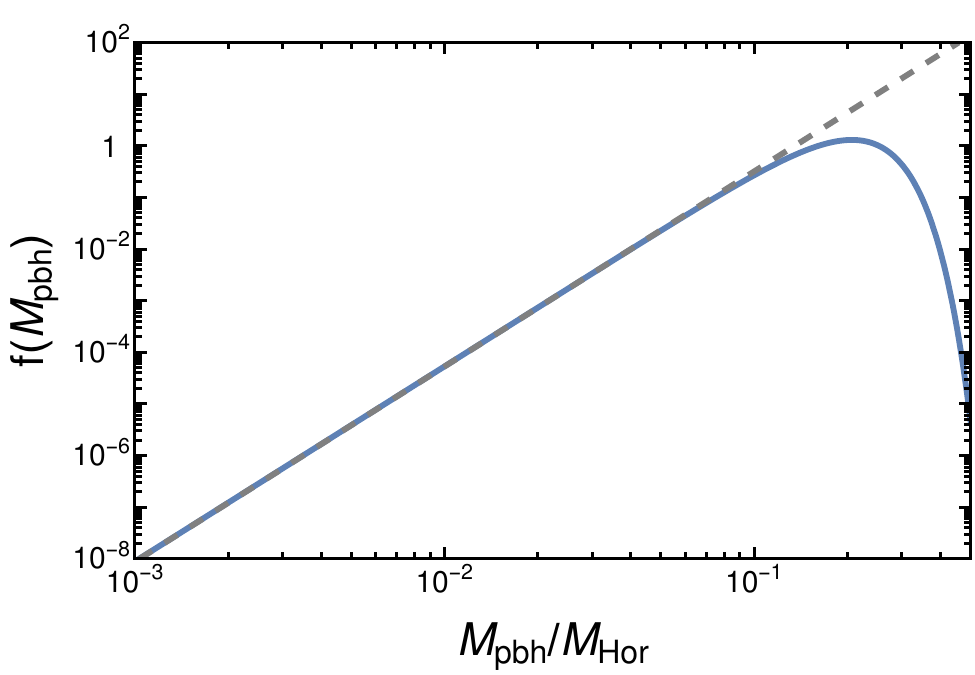}
\end{center}
\caption{\small The PBH spectra for the example parameter point, $v_{\rho} = 10^{3}$ TeV, $y_{Ni} = 0.01$, and $g_{\BL} = 0.2958$, now taking into account that $\delta_{\rm max}$ is a function of $T_{i}$, and assuming the critical collapse scaling with $\mathcal{K}=1$. Upon integration we find $\langle M_{\PBH} \rangle \simeq 5.4 \times 10^{-12} \; M_{\odot}$ and an integrated total $f_{\PBH} \simeq 1.05$, compared with $M_{\PBH} = 4.5 \times 10^{-12} \; M_{\odot}$ and $f_{\PBH} \simeq 1.02$ using the monochromatic approximation. The grey dashed lines show the expected IR scalings, $\Psi(M_{\PBH}) \propto M_{\PBH}^{1/\gamma_c}$ and $f_{\PBH}(M_{\PBH}) \propto M_{\PBH}^{1+1/\gamma_c}$ respectively~\cite{Franciolini:2021nvv}, coming from values of $T_{i}$ close to which $\delta_{\rm max} = 0.45$. These IR scalings can be understood from Eqs.~\eqref{eq:nonmonochrommspectrum1} and \eqref{eq:nonmonochromfspectrum2}.  }
\label{fig:pbhspectrum}
\end{figure}

\begin{figure}[h!]
\begin{center} 
 \includegraphics[width=260pt]{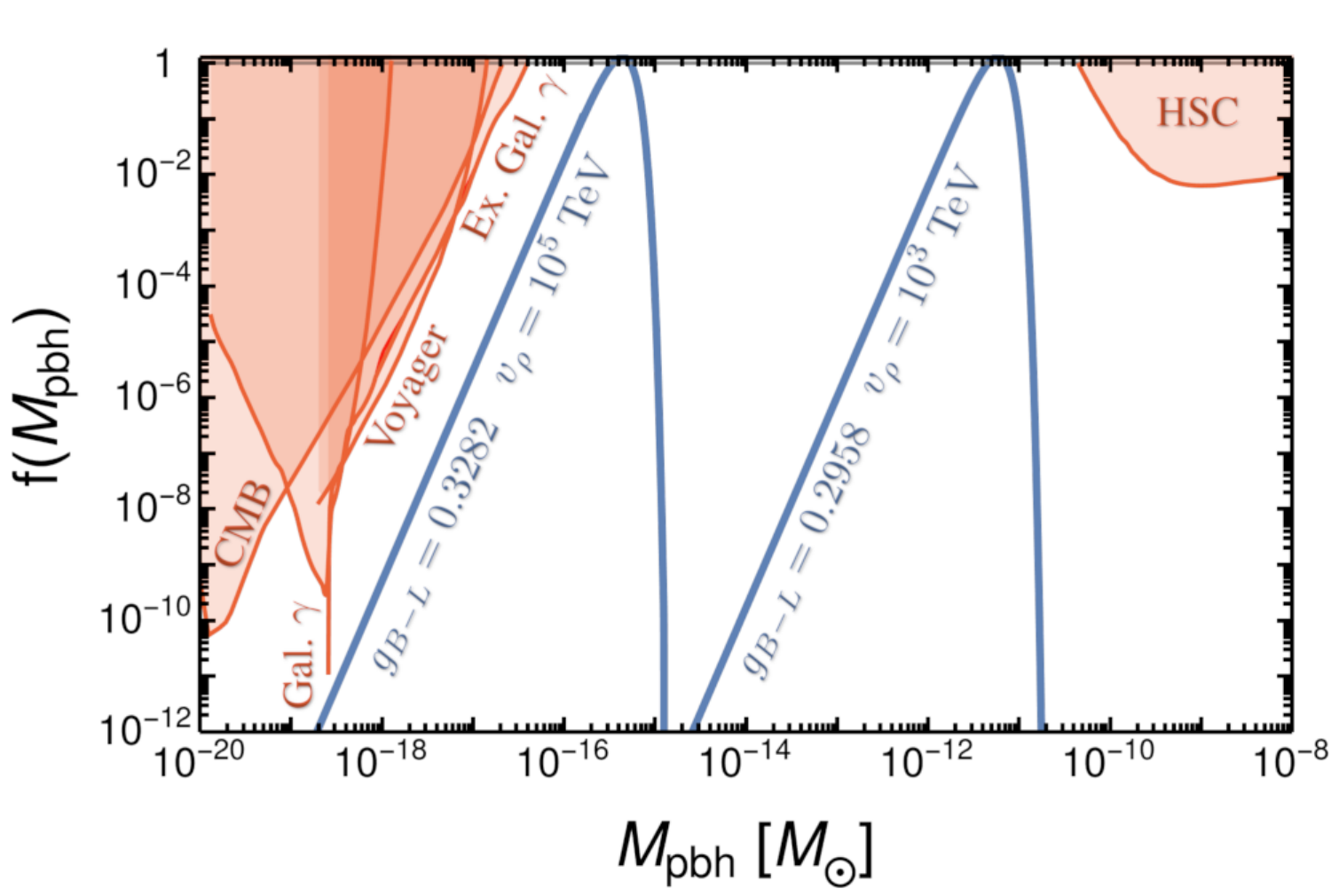}
\end{center}
\caption{\small The PBH spectra for two example parameter points, both with $y_{Ni}=0.01$ and $\mathcal{K}=1$, compared to limits on monochromatic spectra from the CMB~\cite{Acharya:2020jbv,Chluba:2020oip}, extra galactic background light~\cite{Carr:2009jm}, galactic gamma ray background~\cite{Carr:2020gox}, Voyager $e^{\pm}$~\cite{Boudaud:2018hqb}, and Subaru Hyper Suprime-Cam (HSC) lensing~\cite{Niikura:2017zjd}. The GWs from the same two parameter points are shown in Fig.~\ref{fig:GWSPECexample}.}
\label{fig:pbhspectrum2}
\end{figure}

We now return to our example parameter point used in Fig.~\ref{fig:densities} and calculate the PBH mass spectra assuming the critical collapse phenomenon holds. The resulting PBH spectra are shown in Fig.~\ref{fig:pbhspectrum}. Clearly the spectra are strongly peaked at $M_{\PBH} \approx c_{\rm sound}^{3} M_{\rm Hor} \simeq 0.2 M_{\rm Hor}$, the value assumed for the monochromatic approximation. Furthermore, the total integrated PBH density normalized to the observed DM density remains largely unchanged. Thus the monochromatic approximation is justified. We can also compare the resulting spectra to existing limits as done in Fig.~\ref{fig:pbhspectrum2}, but only to gain a rough idea, as the limits are strictly only valid in the monochromatic approximation. Still it is reassuring that the spectra are strongly enough peaked that the IR tails do not intersect with the low mass constraints, as long as $\langle M_{\PBH} \rangle \gtrsim \mathcal{O}(10^{-16}) \; M_{\odot}$. The UV cut-off is much steeper and so even less of a concern.

Even assuming the critical collapse relation holds for the late patch mechanism, the above is still not a complete estimate of the full PBH spectrum. This is because: (i) we have ignored the possibility of PBHs being formed at slightly different times with different $M_{\rm Hor}$, and (ii) the collapsing volume is not limited to Eq.~\eqref{eq:volfactor}, but may take a whole range of values (for us, horizon and superhorizon at the time of the PT, but one may also imagine that sub-horizon patches with larger densities can also collapse, as in Ref.~\cite{Lewicki:2023ioy}).  Given the theoretical uncertainties involved in applying the critical collapse to the PT scenario, however, together with the computational time required for the numerical evaluation, we have not yet evaluated this full expression. Nevertheless, the peaked nature of the PBH spectrum is expected to hold also in this more general case, because the probability of obtaining larger $V_{\rm coll}$ will also fall rapidly.

\medskip
\small

\bibliographystyle{JHEP}
\bibliography{BibPTPBH}
\end{document}